\documentclass{article}
\usepackage{siunitx}
\usepackage{amsmath,amssymb}
\usepackage{geometry}
\usepackage{placeins}
\usepackage{hyperref} % hyperlinks in pdf, very useful
\usepackage{doi}
\usepackage{multirow}
\usepackage{graphicx}
\usepackage[caption=false]{subfig}
\title{A modelling and computational study of the frustration index in signed networks}

\author{Samin Aref$^1$ $^2$ \and Andrew J.~Mason$^3$ \and Mark C.~Wilson$^2$}

\date{%
	$^1$Max Planck Institute for Demographic Research\\ Konrad-Zuse-Str. 1, Rostock, 18057, Germany\\%
	$^2$School of Computer Science, University of Auckland\\ Auckland, Private Bag 92019, New Zealand\\%
	$^3$Department of Engineering Science, University of Auckland\\ Auckland, Private Bag 92019, New Zealand\\%
	{sare618@aucklanduni.ac.nz}\\[2ex]%
	\today
}
\begin{document}
	\maketitle

%\contrib[\authfn{1}]{Equally contributing authors.}

% Include full affiliation details for all authors

%\presentadd[\authfn{2}]{Department, Institution, City, State or Province, Postal Code, Country}

%\fundinginfo{There is no funding sources to be reported for this study.}

% Include the name of the author that should appear in the running header
%\runningauthor{Samin Aref et al.}

\begin{abstract}
Computing the frustration index of a signed graph is a key step toward solving problems in many fields including social networks, political science, physics, chemistry, and biology. The frustration index determines the distance of a network from a state of total structural balance. Although the definition of the frustration index goes back to the 1950's, its exact algorithmic computation, which is closely related to classic NP-hard graph problems, has only become a focus in recent years. We develop three new binary linear programming models to compute the frustration index exactly and efficiently as the solution to a global optimisation problem. Solving the models with prioritised branching and valid inequalities in Gurobi, we can compute the frustration index of real signed networks with over 15000 edges in less than a minute on inexpensive hardware. We provide extensive performance analysis for both random and real signed networks and show that our models outperform all existing approaches by large factors. Based on solve time, algorithm output, and effective branching factor we highlight the superiority of our models to both exact and heuristic methods in the literature.\\
%\footnote{The current authors have published a book chapter under the title ``Computing the Line Index of Balance Using Integer Programming Optimisation''\cite{aref2017computing}. This paper is a continuation of the same line of research with a focus on developing new models to reduce solve time by large factors and facilitate processing larger instances.}

% Please include a maximum of seven keywords

{The reference to this article should be made as follows: {\scshape Aref, S., Mason, M.~J., Wilson, M.~C.}
	\newblock A modelling and computational study of the frustration index in signed networks.
	\newblock {\em Networks}, (Forthcoming), 
	\newblock doi: \href{https://onlinelibrary.wiley.com/journal/10970037}{10.1002/net.XXXXX}.}\\

Old title: An exact method for computing the frustration index in signed networks using binary programming

\end{abstract}

\clearpage

\section{Introduction} \label{3s:intro}
Local ties between entities lead to global structures in networks. Ties can be formed as a result of interactions and individual preferences of the entities in the network. The dual nature of interactions in various contexts means that the ties may form in two opposite types, namely positive ties and negative ties. In a social context, this is interpreted as friendship versus enmity or trust versus distrust between people. The term \textit{signed network} embodies a multitude of concepts involving relationships characterisable by ties with plus and minus signs. \textit{Signed graphs} are used to model such networks where edges have positive and negative signs. \textit{Structural balance} in signed graphs is a macro-scale structural property that has become a focus in network science. Balance theory was the first attempt to understand the sources of tensions and conflicts in groups of people with signed ties \cite{heider_social_1944}. According to balance theory, some structural configurations of people with signed ties lead to social tension and therefore are not balanced.

In network context, if the vertex set of a signed network can be partitioned into $k \leq 2$ subsets such that each negative edge joins vertices belonging to different subsets, it is called a \textit{balanced} network \cite{cartwright_structural_1956}. Using graph-theoretic concepts, Cartwright and Harary identified cycles of the graph %(closed-walks with distinct nodes)
as the origins of tension, in particular cycles containing an odd number of negative edges \cite{cartwright_structural_1956}. By definition, signed graphs in which no such cycles are present satisfy the property of structural balance. For graphs that are not totally balanced, a distance from total balance (a measure of partial balance \cite{aref2015measuring}) can be computed. Among various measures is the \textit{frustration index} that indicates the minimum number of edges whose removal (or equivalently, negation) results in balance \cite{abelson_symbolic_1958,harary_measurement_1959,zaslavsky_balanced_1987}. In what follows, we discuss previous works related to the frustration index (also called the \textit{line index of balance} \cite{harary_measurement_1959}). We use both names, line index of balance and frustration index, interchangeably in this paper. %We refer to the time taken by an algorithm on an instance of a problem as \textit{solve time} (which is equal to \textit{solve time} if the instance is solved).

\subsection{Motivation}

In the past few decades, different measures of balance \cite{cartwright_structural_1956,norman_derivation_1972,terzi_spectral_2011,kunegis_applications_2014,estrada_walk-based_2014} have been suggested and deployed to analyse balance in real-world signed networks resulting in conflicting observations \cite{leskovec_signed_2010, facchetti_computing_2011, estrada_walk-based_2014}. Measures based on cycles \cite{cartwright_structural_1956,norman_derivation_1972}, triangles \cite{terzi_spectral_2011,kunegis_applications_2014}, and closed-walks \cite{estrada_walk-based_2014} are not generally consistent and do not satisfy key axiomatic properties \cite{aref2015measuring}. Among all the measures, a normalised version of the frustration index is shown to satisfy many basic axioms \cite{aref2015measuring}. This measure provides a clear understanding of the transition to balance in terms of the number of edges to be modified to reduce the tension, as opposed to graph cycles that were first suggested as origins of tension in unbalanced networks \cite{cartwright_structural_1956}.

The frustration index is a key to frequently stated problems in many different fields of research \cite{iacono_determining_2010,harary_signed_2002,kasteleyn_dimer_1963,patrick_doreian_structural_2015,doslic_computing_2007}. In biological networks, optimal decomposition of a network into monotone subsystems is made possible by computing the frustration index \cite{iacono_determining_2010}. In finance, performance of a portfolio can be linked to the balance of its underlying signed graph \cite{harary_signed_2002}. In physics, the frustration index provides the minimum energy state in models of atomic magnets known as Ising models \cite{kasteleyn_dimer_1963}. In political science \cite{aref-neal2019legislative} and international relations \cite{patrick_doreian_structural_2015}, networks can be partitioned into cohesive clusters using the line index of balance. In chemistry, bipartite edge frustration has applications to the stability of fullerene, a carbon allotrope \cite{doslic_computing_2007}. For discussions on applications of the frustration index, one may refer to \cite{aref2017balance}.

\subsection{Complexity}
Computing the frustration index is related to the well-known unsigned graph optimisation problem EDGE-BIPARTIZATION, which requires minimisation of the number of edges whose deletion makes the graph bipartite. Given an instance of the latter problem, by declaring each edge to be negative we convert it to the problem of computing the frustration index. Since EDGE-BIPARTIZATION is known to be NP-hard \cite{yannakakis1981edge}, so is computing the frustration index. In the converse direction there is a reduction of the frustration index problem to EDGE-BIPARTIZATION which increases the number of edges by a factor of at most $2$ \cite{huffner_separator-based_2010}. If the reduction preserves planarity, the frustration index can be computed in polynomial time for such planar graphs \cite{grotschel1981weakly}, which is equivalent to the ground state calculation of a two-dimensional spin glass model with no periodic boundary conditions and no magnetic field \cite{DeSimone1995,hartmann2016revisiting}. 

The classic graph optimisation problem MAXCUT is also a special case of the frustration index problem, as can be seen by assigning all edges to be negative (an edge is frustrated if and only if it does not cross the cut). 

\subsection{Approximation}
In general graphs, the frustration index is even NP-hard to approximate within any constant factor (assuming Khot's Unique Games Conjecture \cite{khot2002power}) \cite{huffner_separator-based_2010}. That is, for each $C>0$, the problem of finding an approximation to the frustration index that is guaranteed to be within a factor of $C$ is believed to be NP-hard.

%The computational complexity of the problem might have played a role in the lack of systematic investigation of computing the frustration index while there are many studies on approximating it.
%Efficient approximations have long existed for MAXCUT including Goemans and Williamson's approximation featuring an approximation guarantee of 0.878 \cite{goemans_improved_1995}. The semidefinite programming algorithm of Goemans and Williamson leads to the same approximation guarantee for the frustration index \cite{thagard1998coherence, dasgupta_algorithmic_2007}. %%SA This is probably wrong because we cannot approximate it to a constant factor
The frustration index can be approximated to a factor of $\mathcal{O}(\sqrt{\log n})$ \cite{agarwal2005log} or $\mathcal{O}(k \log k)$ \cite{avidor2007multi} where $n$ is the number of vertices and $k$ is the frustration index. Coleman et al.\ provide a review on the performance of several approximation algorithms of the frustration index \cite{coleman2008local}.

\subsection{Heuristics and local optimisation}
Doreian and Mrvar have reported numerical values as the line index and suggest that determining this index is in general a polynomial-time hard problem \cite{patrick_doreian_structural_2015}. However, their algorithm does not provide optimal solutions and the results are not equal to the line index of balance \cite{aref2017computing}.
Data-reduction schemes \cite{huffner_separator-based_2010} and ground state search heuristics \cite{iacono_determining_2010} are used to obtain estimates for the frustration index. 
%Iacono et al.\ showed that the frustration index equals the minimum number of fundamental negative cycles induced over all spanning trees of the graph \cite{iacono_determining_2010}. Originally discussed in the biology context, the terminology used in \cite{iacono_determining_2010} is different where \textit{monotonocity} and \textit{consistency deficit} are the equivalents for balance and frustration.
Facchetti, Iacono, and Altafini suggested a non-linear energy function minimisation model for finding the frustration index \cite{facchetti_computing_2011}. Their model was solved using various techniques \cite{iacono_determining_2010, esmailian_mesoscopic_2014, ma_memetic_2015, ma_decomposition-based_2017}. Using the ground state search heuristic algorithms \cite{iacono_determining_2010}, the frustration index is estimated in biological networks with $n \approx 1.5\times10^3$ \cite{iacono_determining_2010} and social networks with $n \approx 10^5$ \cite{facchetti_computing_2011, facchetti2012exploring}. 

\subsection{Exact computation}
Using a parametrised algorithmics approach, H\"{u}ffner, Betzler, and Niedermeier show that the frustration index (under a different name) is \textit{fixed parameter tractable} and can be computed in $\mathcal{O}(2^k m^2)$ \cite{huffner_separator-based_2010}, where $m$ is the number of edges and $k$ is the fixed parameter (the frustration index). %The values of $k$ we have observed in signed graphs inferred from the literature make this approach impractical (see Subsection~\ref{3s:real} for numerical results on real networks).
Binary (quadratic and linear) programming models were recently suggested as methods for computing the exact value of the frustration index \cite{aref2017computing} capable of processing graphs with $m \approx 10^3$ edges.

%Their suggested heuristic is reported to solve networks with up to $n \leq 10^5$ within $99\%$ of optimality. However, not only their main theorem (Theorem 1 in \cite{esmailian_mesoscopic_2014}) is incorrect, but Mendon{\c{c}}a et al.\ has also cast doubt on their main conclusion regarding the role of negative ties in signed graphs \cite{mendoncca2015relevance}.

\subsection{Related works on a similar problem}\label{3s:related}
Despite the lack of exact computational methods for the frustration index, a closely related and more general problem in signed networks has been investigated comprehensively. According to Davis's definition of \textit{generalised balance}, a signed network is \textit{weakly balanced} ($k$-balanced) if and only if its vertex set can be partitioned into $k$ subsets such that each negative edge joins vertices belonging to different subsets \cite{Davis}. The problem of finding the minimum number of frustrated edges for general $k$ (an arbitrary number of subsets) is referred to as the \textit{Correlation Clustering} problem \cite{bansal2004correlation}.

For every fixed $k$, there is a polynomial-time approximation scheme for the correlation clustering problem \cite{Giotis}. For arbitrary $k$, exact \cite{brusco_k-balance_2010, figueiredo2013mixed} and heuristic methods \cite{drummond2013efficient,levorato2015ils,levorato2017evaluating} are developed based on a mixed integer programming model \cite{demaine2006correlation}. Denoting the order of a graph by $n$, exact algorithms fail for $n>21$ \cite{brusco_k-balance_2010} and $n>40$ \cite{figueiredo2013mixed}, while greedy algorithms \cite{drummond2013efficient} and local search heuristics \cite{levorato2015ils} are used for larger instances with $n \approx 10^3$ and $n \approx 10^4$ respectively.

After extending the non-linear energy minimisation model suggested by Facchetti et al.\ \cite{facchetti_computing_2011} to generalised balance, Ma et al.\ has experimented on the correlation clustering problem in networks with $n \approx 10^5$ using various heuristics \cite{ma_memetic_2015, ma_decomposition-based_2017}. Esmailian et al.\ have also extended the work of Facchetti et al.\ \cite{facchetti_computing_2011} focusing on the role of negative ties in signed graph clustering \cite{esmailian_mesoscopic_2014, esmailian2015community}.

\subsection*{Our contribution} \label{3ss:contrib}
The principal focus of this research study is to provide further insight into computing the frustration index by developing efficient computational methods outperforming previous methods by large factors. We systematically investigate several formulations for exact computation of the frustration index and compare them based on solve time as well as other performance measures.

The advantage of formulating the problem as an optimisation model is not only exploring the details involved in a fundamental NP-hard problem, but also making use of powerful mathematical programming solvers like Gurobi \cite{gurobi} to solve the NP-hard problem exactly and efficiently. We provide numerical results on a variety of undirected signed networks, both randomly generated and inferred from well-known data sets (including real signed networks with over 15000 edges).

A recent study by the current authors has investigated computing the frustration index in smaller scales using quadratic and linear optimisation models \cite{aref2017computing}. The linear model is used for computing the frustration index in several small random and real networks with up to 3200 edges. We improve the contributions of \cite{aref2017computing} by providing three new binary linear formulations which not only outperform the models in \cite{aref2017computing} by large factors, but also facilitate a more direct and intuitive interpretation. We discuss more efficient speed-up techniques that require substantially fewer additional constraints compared to \cite{aref2017computing}. This allows Gurobi's branch and bound algorithm to start with a better root node solution and explore considerably (several orders of magnitude) fewer nodes leading to a substantially shorter solve time. Moreover, our new models handle order-of-magnitude larger instances that are not solvable by the models in \cite{aref2017computing}. %In this study, we provide a more straight forward interpretation of the optimisation models. %obviating the need to discuss some convoluted signed graph concepts. 
We provide in-depth performance analysis using extensive numerical results showing the solve times of our worst-performing model to be $2 - 9$ times faster than the best-performing model in \cite{aref2017computing}.

This paper begins by laying out the theoretical dimensions of the research in Section~\ref{3s:prelim}. Linear programming models are formulated in Section~\ref{3s:model}. Section~\ref{3s:speed} provides different techniques to improve the formulations and reduce solve time. The numerical results on the models' performance are presented in Section~\ref{3s:results}. Section~\ref{3s:evaluate} provides comparison against the literature using both random and real networks. %Recent developments on a closely related problem are discussed in Section~\ref{3s:related} followed by 
Other formulations and extensions to the models are provided in Section~\ref{3s:future} followed by Section~\ref{3s:conclu} which sums up the research highlights. 

\section{Preliminaries} \label{3s:prelim}

We recall some standard definitions.

\subsection{Basic notation} \label{3s:problem}
We consider undirected signed networks $G = (V,E,\sigma)$. The ordered set of nodes is denoted by $V=\{1,2,\dots,n\}$, with $|V| = n$. The set $E$ of edges is partitioned into the set of positive edges $E^+$ and the set of negative edges $E^-$ with $|E^-|=m^-$, $|E^+|=m^+$, and $|E|=m=m^- + m^+$. The sign function is denoted by $\sigma: E\rightarrow\{-1,+1\}$.

We represent the $m$ undirected edges in $G$ as ordered pairs of vertices $E = \{e_1, e_2, ..., e_m\} \subseteq \{ (i,j) \mid i,j \in V , i<j \}$, where a single edge $e_k$ between nodes $i$ and $j$, $i<j$, is denoted by $e_k=(i,j) , i<j$. We denote the graph density by $\rho= 2m/(n(n-1))$. The entries, $a{_i}{_j}$, of the \emph{signed adjacency matrix}, \textbf{A}, are defined in \eqref{3eq1}. 
\begin{equation}\label{3eq1}
a{_i}{_j} =
\left\{
\begin{array}{ll}
\sigma_{(i,j)} & \mbox{if } (i,j) \in E \\
\sigma_{(j,i)} & \mbox{if } (j,i) \in E \\
0 & \text{otherwise}
\end{array}
\right.
\end{equation}

The number of edges incident to the node $i \in V$ represents the degree of node $i$ and is denoted by $d {(i)}$. A directed cycle (for simplicity \emph{cycle}) of length $k$ in $G$ is a sequence of nodes $v_0,v_1,...,v_{k-1},v_k=v_0$ such that for each $i=1,2,...,k$ there is an edge from $v_{i-1}$ to $v_i$. The \emph{sign} of a cycle is the product of the signs of its edges. A cycle with negative sign is unbalanced. A balanced cycle is one with positive sign. A balanced graph is one with no negative cycles.

\subsection{Node colouring and frustration count}\label{3ss:frustrationcount}

\begin{figure}[ht]
	\subfloat[An example graph with four nodes, two positive edges, and three negative edges]{\includegraphics[height=1.2in]{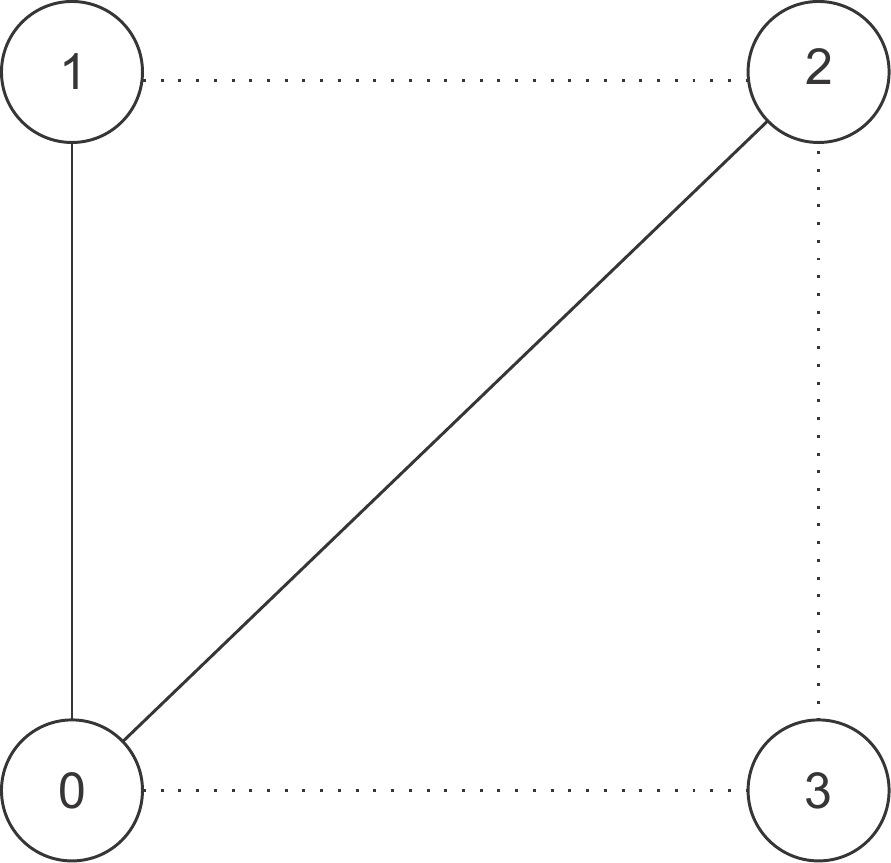}%
		\label{3fig1a}}
	\hfil
	\subfloat[An arbitrary node colouring resulting in two frustrated edges (0,2), (2,3)]{\includegraphics[height=1.2in]{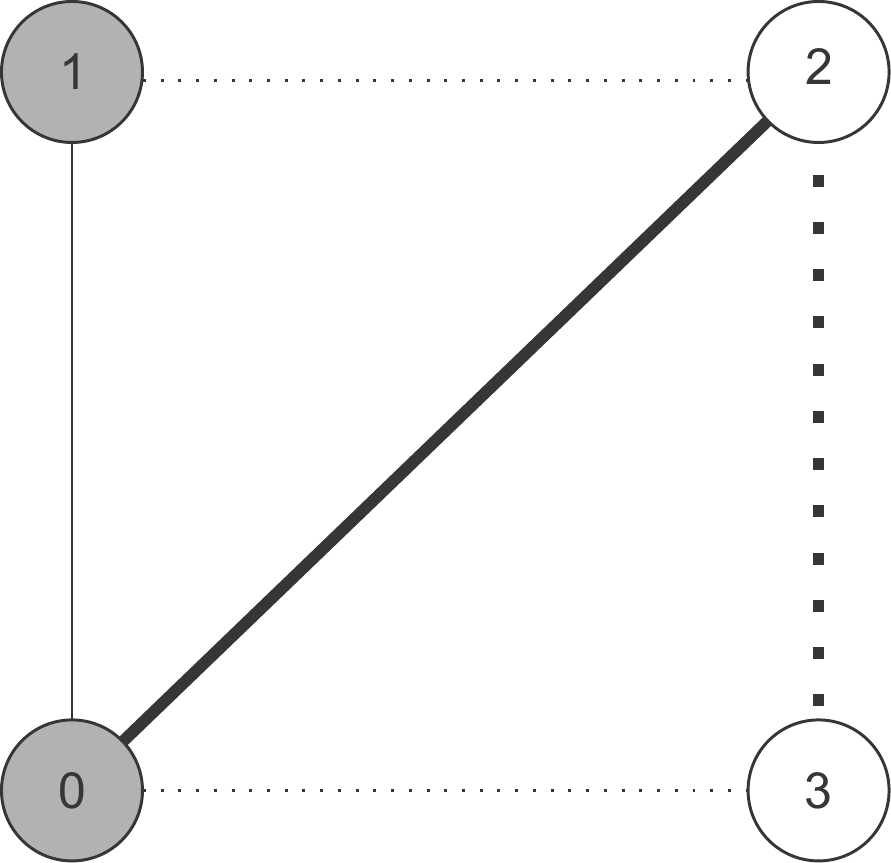}%
		\label{3fig1b}} 
	\hfil
	\subfloat[Another node colouring resulting in one frustrated edge (1,2)]{\includegraphics[height=1.2in]{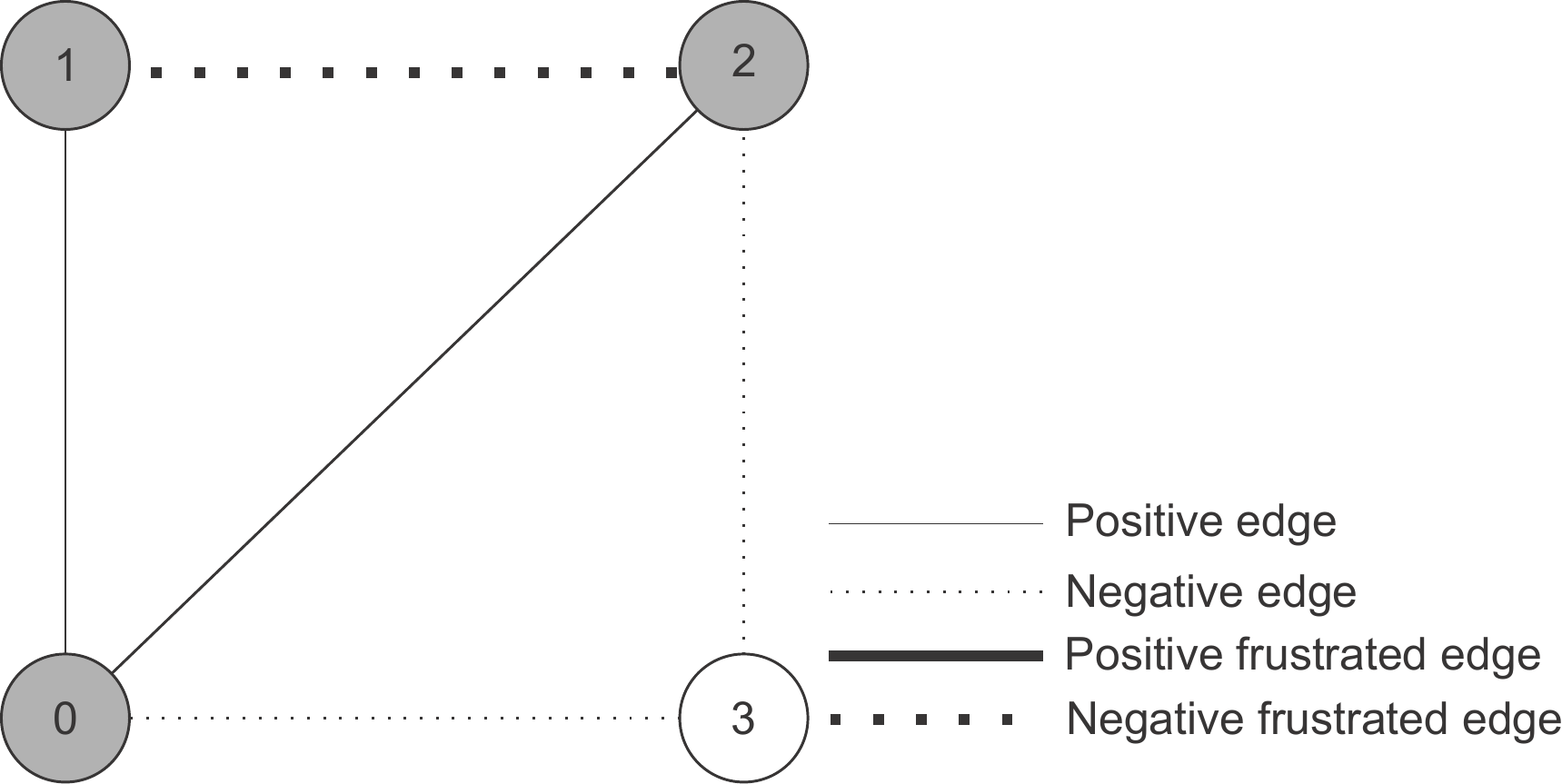}%
		\label{3fig1c}}
	\caption{Node colourings and the respective frustrated edges for an example signed graph}% The figure is produced using Adobe Illustrator.}
	\label{3fig1}
\end{figure}

\textit{Satisfied} and \textit{frustrated} edges are defined based on colourings of the nodes. Colouring each node with black or white, a frustrated (satisfied) edge $(i,j)$ is either a positive (negative) edge with different colours on the endpoints $i,j$ or a negative (positive) edge with the same colours on the endpoints $i,j$. Subfigure \ref{3fig1a} illustrates an example signed graph in which positive and negative edges are represented by solid lines and dotted lines respectively. Subfigures \ref{3fig1b} and \ref{3fig1c} illustrate node colourings and their impacts on the frustrated edges that are represented by thick lines.

%\textbf{Definition.}
%For any signed graph $G=(V, E, \sigma)$, we can partition $V$ into two sets, denoted $X \subseteq V$ and $\overline{X}=V \backslash X$. We call $X$ the colouring set and we think of this partitioning as specifying a colouring of the nodes, where each node $i \in X$ is coloured black, and $i \in \overline{X}$ is coloured white. We let $x_i$ denote the colour of node $i \in V$ under $X$, where $x_i=1$ if $i \in X$ and $x_i=0$ otherwise. 
Let $X \subseteq V$ be a subset of vertices. This defines a partition $(X, V \setminus X)$ of $V$. We call $X$ a colouring set. 
%\textbf{Definition.}
Let binary variable $x_i$ denote the colour of node $i \in V$ under colouring set $X$. We consider $x_i=1$ if $i \in X$ (black node) and $x_i=0$ if $i \in V \setminus X$ (white node). %We say that each node $i \in X$ is black and each node $i \in V \backslash X$ is white. 
%We say that an edge $(i,j)$ is frustrated under $X$ if either edge $(i,j)$ is a positive edge (i.e.\ $(i,j) \in E^+$) but nodes $i$ and $j$ have different colours ($x_i \ne x_j$), or edge $(i,j)$ is a negative edge (i.e.\ $(i,j) \in E^-$) but nodes $i$ and $j$ share the same colour ($x_i = x_j$). 

\textbf{Definition.}
We define the {\em frustration count} of signed graph $G$ under colouring $X$ as $f_G(X) := \sum_{(i,j) \in E} f_{ij}(X)$
where $f_{ij}(X)$ is the frustration state of edge $(i,j)$, given by
\begin{equation} \label{3eq2}
f_{ij}(X)=
\begin{cases}
0, & \text{if}\ x_i = x_j \text{ and } (i,j) \in E^+ \\
1, & \text{if}\ x_i = x_j \text{ and } (i,j) \in E^- \\
0, & \text{if}\ x_i \ne x_j \text{ and } (i,j) \in E^- \\
1, & \text{if}\ x_i \ne x_j \text{ and } (i,j) \in E^+ .\\
\end{cases}
\end{equation}

The optimisation problem consists in finding a subset $X^* \subseteq V$ of $G$ that minimises the frustration count $f_G(X)$, i.e., solving Eq.\ \eqref{3eq3}. The globally optimal solution to this problem gives the frustration index $L(G)$ of signed graph $G$.
\begin{equation} \label{3eq3}
L(G) = \min_{X \subseteq V}f_G(X)\
\end{equation}

It follows that $f_G(X)$ gives an upper bound on $L(G)$ for any $X \subseteq V$. Note that the colouring in Subfigure \ref{3fig1b} does not minimise $f_G(X)$, while in Subfigure \ref{3fig1c} $f_G(X)$ is minimum.

\section{Binary linear programming formulations} \label{3s:model}
In this section, we introduce three binary linear models in \eqref{3eq3.5} -- \eqref{3eq5} to minimise the frustration count as the objective function. There are various ways to form the frustration count using variables defined over graph nodes and edges which lead to various mathematical programming models that we discuss in this section. 

\subsection{The AND model}
We start with an objective function to minimise the frustration count. Note that the frustration state of a positive edge $(i,j)$ can be represented by $f_{ij}=x_i +x_j - 2x_ix_j$ $\forall (i,j) \in E^+$ using the two binary variables $x_i , x_j\in\{0,1\}$ for the endpoint colours. For a negative edge, we have $f_{ij}= 1- (x_i +x_j - 2x_ix_j)$ $\forall (i,j) \in E^-$.

The term $x_i x_j$ can be replaced by binary variable $x_{ij}=x_i x_j$ for each edge $(i,j)$ that take value 1 whenever $\text{AND}_{(x_i,x_j)}=1$ (both endpoints are coloured black) and 0 otherwise. This gives our first binary linear model in \eqref{3eq3.5} that calculates the frustration index in the minimisation objective function.

The optimal solution represents a subset $X^* \subseteq V$ of $G$ that minimises the frustration count. The optimal value of the objective function in Eq.\ \eqref{3eq3.5} is denoted by $Z^*$ which represents the frustration index.

The dependencies between the $x_{ij}$ and $x_{i},x_{j}$ values are taken into account using standard AND constraints. The AND model has $n+m$ variables and $2m^+ + m^-$ constraints. Note that $x_{ij}$ variables are dependent variables because of the constraints and the minimisation objective function. Therefore, we may drop the integrality constraint of the $x_{ij}$ variables and consider them as continuous variables in the unit interval, $x_{ij} \in [0,1]$. The next subsection discusses an alternative binary linear model for computing the frustration index.

\begin{equation}\label{3eq3.5}
\begin{split}
\min_{x_i: i \in V, x_{ij}: (i,j) \in E} Z &= \sum\limits_{(i,j) \in E^+} x_{i} + x_{j} - 2x_{ij}  + \sum\limits_{(i,j) \in E^-} 1 - (x_{i} + x_{j} - 2x_{ij})\\
\text{s.t.} \quad
x_{ij} &\leq x_{i} \quad \forall (i,j) \in E^+ \\
x_{ij} &\leq x_{j} \quad \forall (i,j) \in E^+ \\
x_{ij} &\geq x_{i}+x_{j}-1 \quad \forall (i,j) \in E^- \\
x_{i} &\in \{0,1\} \quad  \forall i \in V \\
x_{ij} &\in \{0,1\} \quad \forall (i,j) \in E 
\end{split}
\end{equation}

\subsection{The XOR model}
%Minimising the frustration count can be directly formulated as a binary linear model. 
The XOR model is designed to directly count the frustrated edges using binary variables $f_{ij}\in\{0,1\}, \forall (i,j) \in E$. As before, we use $x_i\in\{0,1\}, \forall i \in V$ to denote the colour of node $i$. This model is formulated by observing that the frustration state of a positive edge $(i,j) \in E^+$ is given by $f_{ij}=\text{XOR}_{(x_i,x_j)}$. Similarly for $(i,j) \in E^-$, we have $f_{ij}=1- \text{XOR}_{(x_i,x_j)}$. Therefore, the minimum frustration count under all node colourings is obtained by solving \eqref{3eq4}.
\begin{equation}\label{3eq4}
\begin{split}
\min_{x_i: i \in V, f_{ij}: (i,j) \in E} Z &= \sum\limits_{(i,j) \in E}  f_{ij}  \\
\text{s.t.} \quad
f_{ij} &\geq x_{i}-x_{j} \quad \forall (i,j) \in E^+ \\
f_{ij} &\geq x_{j}-x_{i} \quad \forall (i,j) \in E^+ \\
f_{ij}  &\ge  x_{i} + x_{j} -1 \quad \forall (i,j) \in E^- \\
f_{ij}  &\ge  1-x_{i} - x_{j}  \quad \forall (i,j) \in E^- \\
x_{i} &\in \{0,1\} \quad  \forall i \in V \\
f_{ij} &\in \{0,1\} \quad \forall (i,j) \in E 
\end{split}
\end{equation}

The dependencies between the $f_{ij}$ and $x_{i},x_{j}$ values are taken into account using two standard XOR constraints per edge. Therefore, the XOR model has $n+m$ variables and $2m$ constraints. Note that $f_{ij}$ variables are dependent variables because of the constraints and the minimisation objective function. Therefore, we may specify $f_{ij}$ variables as continuous variables in the unit interval, $f_{ij} \in [0,1]$. A third linear formulation of the problem is provided in the next subsection.

\subsection{The ABS model}
In this subsection, we propose the ABS model, a binary linear model in which we use two edge variables to represent the frustration state of an edge. We start by observing that for a given node colouring, $|x_i - x_j|=1$ for a positive frustrated edge and $|x_i - x_j|=0$ for a positive satisfied edge $(i,j) \in E^+$. Similarly, $1-|x_i - x_j|=|x_i + x_j -1|$ gives the frustration state of a negative edge $(i,j) \in E^-$.

To linearise the absolute value terms, we introduce additional binary variables $e_{ij}, h_{ij}\in\{0,1\},$ $\forall (i,j) \in E$. We replace $|x_i - x_j|$ with $e_{ij} + h_{ij}$ to represent the frustration state of a positive edge $(i,j) \in E^+$. This requires adding the constraint $x_i - x_j = e_{ij} - h_{ij} \ \forall (i,j) \in E^+ $. Similarly, we replace $|x_i + x_j -1|$ with $e_{ij} + h_{ij}$ to represent the frustration state of a negative edge $(i,j) \in E^-$. Accordingly, we add the constraint $x_i + x_j -1 = e_{ij} - h_{ij} \ \forall (i,j) \in E^- $.

These two replacements allow us to linearise the two absolute value terms and formulate the ABS model in \eqref{3eq5} which has $n+2m$ variables and $m$ constraints. Note that in an optimal solution, variables $e_{ij}$ and $h_{ij}$ both take the value $0$ for a satisfied edge $(i,j) \in E$, whereas for a frustrated edge $(i,j) \in E$ exactly one of the two variables $e_{ij}$ and $h_{ij}$ takes the value $1$. The objective function in \eqref{3eq5} sums the frustration states of all edges and its optimal value equals the frustration index. 
\begin{equation}\label{3eq5}
\begin{split}
\min_{x_i: i \in V, e_{ij},h_{ij}: (i,j) \in E} Z &=\sum\limits_{(i,j) \in E}  e_{ij} + h_{ij} \\
\text{s.t.} \quad 
x_{i} - x_{j} &=  e_{ij} - h_{ij} \quad \forall (i,j) \in E^+ \\
x_{i} + x_{j} -1 &= e_{ij} - h_{ij} \quad \forall (i,j) \in E^- \\
x_{i} &\in \{0,1\} \quad  \forall i \in V \\
e_{ij} &\in \{0,1\} \quad \forall (i,j) \in E \\
h_{ij} &\in \{0,1\} \quad \forall (i,j) \in E 
\end{split} 
\end{equation}
%Therefore, the optimal value of $e_{ij} + h_{ij}$ represents the frustration state of edge $(i,j) \in E$.
%Note that for any choice of $x_i$ and $x_j$, we can satisfy the constraint for edge $(i,j) \in E$ with (amongst other possible solutions) $e_{ij}=0$ and $h_{ij}=0$ or $e_{ij}=0$ and $h_{ij}=1$ or $e_{ij}=1$ and $h_{ij}=0$. Because we minimise $e_{ij} + h_{ij}$, these are the solutions we choose. It is the combination of the objective function and the constraints together that makes the ABS model work. %The conditions observed for positive and negative edges are expressed as linear constraints in \eqref{3eq5}. Therefore, the ABS model has $n+2m$ variables and $m$ constraints.

\subsection{Comparison of the models} \label{3ss:compare}
In this subsection we compare the three models introduced above and two of the models suggested in \cite{aref2017computing}, based on the number and type of constraints. Table~\ref{3tab1} summarises the comparison.

\begin{table}[ht]
	%\centering
	\caption{Comparison of optimisation models developed for computing the frustration index}
	\label{3tab1}
	\begin{tabular}{lp{2cm}p{2.5cm}lll} \hline
		& Aref et~al. UBQP \cite{aref2017computing} & Aref et~al. \quad \quad binary linear \cite{aref2017computing}  & {AND \eqref{3eq3.5}} & {XOR \eqref{3eq4}} & {ABS \eqref{3eq5} } \\ \hline
		Variables                    & $n$   & $n+m$    & $n+m$             & $n+m$             & $n+2m$          \\
		Constraints                  & $0$    & $m^+ + m^-$   & $2m^+ + m^-$               & $2m^+ + 2m^-$              & $m^+ + m^-$             \\
		Constraint type     & -       & linear  & linear            & linear            & linear      \\
		Objective            & quadratic & linear  & linear            & linear            & linear         \\ \hline
	\end{tabular}
\end{table}

In optimal solutions of our three suggested models, the frustration state of edge $(i,j)$ is represented by the corresponding term for edge $(i,j)$ in the objective function. This leads to Eq.\ \eqref{3eq6} which makes a connection between the optimal values of the decision variables in the three models.
\begin{equation}\label{3eq6}
	f_{ij} = e_{ij} + h_{ij} = (1-a_{ij})/2 + a_{ij}(x_i +x_j - 2x_{ij}) 
\end{equation}

Note that not only does the number of constraints scale linearly with graph size, each constraint involves at most 4 variables. Thus the worst-case space usage for solving these models is $\mathcal{O}(n^2)$. The three linear models perform differently in terms of solve time and the number of branch and bound (B\&B) nodes required to solve a given instance. 

Solving large-scale binary programming models is not easy in general \cite{Bilitzky2005} and therefore there is a limit to the size of the largest graph whose frustration index can be computed in a given time. In the next section, we discuss some techniques for improving the performance of Gurobi in solving our suggested binary linear models.

\section{Speed-up techniques}\label{3s:speed}

In this section we discuss techniques to speed up the branch and bound algorithm for solving the binary linear models described in the previous section. %To improve the branch and bound, one may use certain conditions based on the structure of the binary programming problem \cite{Bilitzky2005}.
The branch and bound algorithm can be provided with a list of prioritised variables for branching which may speed up the solver if branching on these variables leads more quickly to integer solutions.

Another technique often deployed in solving Integer Programming (IP) models is using \textit{valid inequalities} which we discuss briefly. %They both are well-known techniques in solving integer linear programming models using the branch and bound algorithm.
Two key features of valid inequalities is that they are satisfied by the optimal integer solutions (validity), but are violated by undesired feasible solutions (usefulness). %Furthermore, we hope to find valid inequalities that strengthen the formulation by rejecting undesired integer solutions.
We implement some valid inequalities as \textit{lazy} constraints which %are kept aside from the original constraints of the model and pulled into the model only if they are violated by a solution \cite{Klotzpractical}. %Such implementation of additional restrictions is referred to as lazy constraints. 
are given to the solver, but only added to the model if they are violated by a solution \cite{Klotzpractical,gurobi}.
Implementing valid inequalities as lazy constraints restricts the model by removing the undesired solutions that violate them. Such valid and useful restrictions reduce solve time \cite{Klotzpractical}.
%Branching is the main process in the branch and bound algorithm to progress in searching the feasible space for integer solutions.
%In this section, we discuss several speed-up techniques and report the improvement they yield in Subsection \ref{3ss:overall}. %The solve time improvement evaluation is based on 100 Erd\H{o}s-R\'{e}nyi graphs with uniformly random parameters from the ranges $40 \leq n \leq 50$, $0 \leq \rho \leq 1$, and $0 \leq m^-/m \leq 1$.

\subsection{Pre-processing data reduction}
Standard graph pre-processing can be used to reduce graph size and order without changing the frustration index. This may reduce solve time in graphs containing nodes of degree $0$ and $1$ (also called isolated and pendant vertices respectively) and nodes whose removal increases the number of connected components (also called articulation points) \cite{huffner_separator-based_2010}. %We implement some of the data-reduction schemes in \cite{huffner_separator-based_2010}. H\"{u}ffner et al.\ suggest different ways to reduce nodes and edges of the graph that are separated by a small set of vertices called a \textit{separator} \cite{huffner_separator-based_2010}.
We have tested iterative reduction of isolated and pendant vertices as well as decomposing graphs by cutting them into smaller subgraphs using articulation points. %These operations are referred to as data reduction using separators of size 0 and 1 in \cite{huffner_separator-based_2010}.
Our experiments show that reducing isolated and pendant vertices does not considerably affect the solve time. Moreover, the scarcity of articulation points in many graphs in which isolated and pendant vertices have been removed, makes decomposition based on articulation points not particularly useful. %However, H\"{u}ffner et al.\ suggest their data-reduction schemes using separators of size up to 3 to be very effective on reducing solve time in their experiments \cite{huffner_separator-based_2010}.

\subsection{Branching priority and fixing a colour} \label{3ss:branch}
%In binary programming models, the root node solution is integral if the constraint matrix is unimodular and the right hand side vector is integral. However, 
%Most practical integer optimisation problems, including the problem under investigation, do not have an integral root node solution \cite{Bilitzky2005} and therefore require an algorithm like branch and bound for finding integral solutions.
%In order to speed up the algorithms, we consider adding an additional constraint to increase the root node objective function value.
We relax the integrality constraints and observe in the Linear Programming relaxation (LP relaxation) of all three models that there always exists a fractional solution of $x_i=0.5, \forall i \in V$ which gives an optimal objective function value of 0. We can increase the root node objective by fixing one node variable to value $1$. Fixing a node variable also breaks the symmetry that exists and allows changing all node colours to give an equivalent solution. This is similar to fixing the \textit{ghost spin} in the ground state calculation of a spin glass model \cite{DeSimone1995} and is also used in \cite{aref2017computing}. 

When the colour of node $k$ is fixed by imposing $x_k=1$, the variables associated with edges incident to node $k$ take value $0.5$ (in the ABS model one of the two variables $e_{ij}$ and $h_{ij}$ take value $0.5$). In all three models, this changes the fractional solution of the LP relaxation from $0$ to $d {(k)}/2$ because all edges incident to node $k$ contribute $0.5$ to the objective function. This observation shows that the best node variable to be fixed is the one associated with the highest degree which allows for an increase of $\max_{i \in V} d {(i)}/2$ in the LP relaxation optimal objective function value. We formulate this as a constraint in \eqref{3eq6.5}.
\begin{equation}\label{3eq6.5}
x_{k} = 1 \quad k= \text{arg} \max_{i \in V} d {(i)}
\end{equation}
In our experiments, we always observed an improvement in the root node objective value when Eq.\ \eqref{3eq6.5} was added, which shows it is useful. We provide more detailed results on the root node objective values for several instances in Section~\ref{3s:evaluate}. 

Based on the same idea, we may modify the branch and bound algorithm so that it branches first on the node with the highest degree. This modification is implemented by specifying a branching priority for the node variables in which variable $x_i$ has a priority given by its degree $d {(i)}$.
%Our experiments on random graphs show that fixing a colour and using prioritised branching lead to 60\%, 88\%, and 72\% reduction in the average solve time of AND, XOR, and ABS models respectively.
\subsection{Unbalanced triangle constraints} \label{3ss:unbalanced}
%The structural properties of the problem allow us to restrict the model by adding valid inequalities as additional constraints \cite{Bilitzky2005}. Structural properties of signed graphs can be used to determine valid inequalities. 
%%SA zaslavsky_balance_2010 is not the right reference. We can prove this:
%We know that changing signs of the frustrated edges leads to balance. We also know that all cycles are positive in a balanced graph. When we change the sign of an even number of edges in a cycle, the sign of cycle does not change. Therefore, in order to make all negative cycle positive, an odd number of edges on each of them must change sign.
We consider one valid inequality for each negative cycle of length 3 (unbalanced triangle) in the graph. Under arbitrary colouring $X$, every negative cycle of the graph contains an odd number of frustrated edges. This means that any colouring of the nodes in an unbalanced triangle must produce at least one frustrated edge. Recalling that under colouring $X$, the variable $f_{ij}$ is 1 if edge $(i,j)$ is frustrated (and 0 otherwise), then for any node triple $(i,j,k)$ defining an unbalanced triangle in $G$, inequality \eqref{3eq7} is valid.
\begin{equation}\label{3eq7}
f_{ij} + f_{ik} + f_{jk} 
\geq 1 \quad \forall (i,j,k) \in T^-
\end{equation}

In \eqref{3eq7}, $T^-=\{(i,j,k)\in V^3 \mid a{_i}{_j} a{_i}{_k} a{_j}{_k} = -1 \}$ denotes the set of node triples that define an unbalanced triangle. The expression in inequality \eqref{3eq7} denotes the sum of frustration states for the three edges $(i,j),(i,k),(j,k)$ making an unbalanced triangle. Note that in order to implement the unbalanced triangle constraints \eqref{3eq7}, $f_{ij}$ must be represented using the decision variables in the particular model. Eq.\ \eqref{3eq6} shows how $f_{ij}$ can be defined in the AND and ABS models.
%Note that, $ \min_{f_{ij}: (i,j) \in E} \sum f_{ij}$ subject to \eqref{eq16} with $m$ binary variables $f_{ij} &\in \{0,1\}$ and $|T^-|$ constraints can be considered as another formulation of the problem.
%The valid inequality in \eqref{3eq7} is useful because the fractional solution $x_i=0.5, \forall i \in V$ (associated with an objective function value of 0) violates it. 
We implement the valid inequality in \eqref{3eq7} using Gurobi's feature for adding lazy constraints and ensure that lazy constraints that cut off the relaxation solution at the root node are also pulled into the model (see \textit{lazy} as a tunable parameter in linear constraint attributes in \cite{gurobi}). %From a solve time perspective, our experiments on random graphs show that implementing this speed-up technique leads to 38\%, 24\%, and 12\% reduction in the average solve time of AND, XOR, and ABS models respectively. 

\subsection{Overall improvement made by the speed-up techniques}\label{3ss:overall}
%In Section~\ref{s:speed}, we discussed the solve time improvement made by the individual implementation of the speed-up techniques on the binary linear models in Section~\ref{s:speed}. %The solve time improvement percentages reported are based on 100 random graphs with $n=30,m=300,m^-=150$.
In this subsection, we report the solve time improvement obtained by implementing the speed-up techniques. The evaluation is based on 100 Erd\H{o}s-R\'{e}nyi graphs with uniformly random parameters from the ranges $40 \leq n \leq 50$, $0 \leq \rho \leq 1$, and $0 \leq m^-/m \leq 1$. The total solve time reduction observed when both speed-up techniques (\ref{3ss:branch} -- \ref{3ss:unbalanced}) are implemented is 67\% for the AND model, 90\% for the XOR model, and 78\% for the ABS model. Table \ref{3tab1.5} shows the solve time improvements made by implementing the speed-up techniques individually and collectively.
\begin{table}[ht]
	%\centering
	\caption{Usefulness of the speed-up techniques based on 100 Erd\H{o}s-R\'{e}nyi graphs}
	\label{3tab1.5}
	\begin{tabular}{llllllll}
		\hline
		\multirow{2}{*}{}          & \multicolumn{3}{l}{Average solve time (s)} &  & \multicolumn{3}{l}{Time improvement (\%)} \\ \cline{2-4} \cline{6-8} 
		& AND          & XOR          & ABS          &  & AND            & XOR            & ABS           \\ \hline
		Without speed-up           & 14.80        & 41.60        & 19.71        &  & -              & -              & -             \\
		With branching priority    & 5.90         & 4.91         & 5.50         &  & 60\%           & 88\%           & 72\%          \\
		With triangle inequalities & 9.21         & 31.72        & 17.26        &  & 38\%           & 24\%           & 12\%          \\
		With both speed-up techniques        & 4.93         & 4.08         & 4.42         &  & 67\%           & 90\%           & 78\%          \\ \hline
	\end{tabular}
\end{table}

\section{Computational performance} \label{3s:results}
In this section, our optimisation models are tested on various random instances using 64-bit Gurobi version 7.5.2 on a desktop computer with an Intel Core i5 7600 @ 3.50 GHz (released in 2017) and 8.00 GB of RAM running 64-bit Microsoft Windows 10. We use \textit{NetworkX} package in Python for generating random graphs. The models were created using Gurobi's Python environment in 64-bit Anaconda3 5.0.1 Jupyter.

\subsection{Comparison of the models' performance} \label{3ss:perform}
In this subsection, we discuss the time performance of Gurobi for solving the extended binary linear models which include both speed-up techniques (\ref{3ss:branch} -- \ref{3ss:unbalanced}).

In order to compare the performance of the three linear models, we consider 12 test cases each containing 10 Barab\'{a}si-Albert random graphs with various combinations of density and proportion of negative edges. The results in Table~\ref{3tab2} show that the three models have relatively similar performance in terms of solve time. 
\begin{table}[ht]
	%\centering
	\caption{Solve time comparison of the three models based on test cases of 10 Barab\'{a}si-Albert graphs}
	\label{3tab2}
	\begin{tabular}{llllllll}
		\hline
		$n$ & $m$ &  $\rho$ & $\frac{m^-}{m}$ & Average $Z^*$ & \multicolumn{3}{c}{Solve time (s) mean  $\pm$  SD}                          \\ \cline{6-8} 
		&                        &                          &                          & \multicolumn{1}{r}{}                       & \multicolumn{1}{c}{AND \eqref{3eq3.5}} & \multicolumn{1}{c}{XOR \eqref{3eq4}} & \multicolumn{1}{c}{ABS \eqref{3eq5}} \\ \hline
		60                   & 539                  & 0.3                     & 0.3                      & 157.4                                      & 1.13  $\pm$ 0.48        & 1.59 $\pm$ 0.3          & 0.84 $\pm$ 0.1          \\
		&                      &                         & 0.5                      & 185.0                                      & 1.48 $\pm$ 0.56         & 2.95 $\pm$ 0.28         & 1.1 $\pm$ 0.19          \\
		&                      &                         & 0.7                      & 172.9                                      & 1.07 $\pm$ 0.41         & 2.55 $\pm$ 0.8          & 0.84 $\pm$ 0.16         \\
		&                      &                         & 1                        & 55.0                                       & 0.04 $\pm$ 0.01         & 0.04 $\pm$ 0.01         & 0.06 $\pm$ 0.02         \\
		& 884                  & 0.5                     & 0.3                      & 262.4                                      & 1.4 $\pm$ 0.16          & 0.45 $\pm$ 0.08         & 0.41 $\pm$ 0.04         \\
		&                      &                         & 0.5                      & 325.8                                      & 37.41 $\pm$ 11.53       & 27.09 $\pm$ 27.09       & 25.15 $\pm$ 8.46        \\
		&                      &                         & 0.7                      & 329.4                                      & 36.73 $\pm$ 8.28        & 39.8 $\pm$ 7.82         & 30.44 $\pm$ 5.73        \\
		&                      &                         & 1                        & 272.4                                      & 1 $\pm$ 0.17            & 0.77 $\pm$ 0.26         & 6.12 $\pm$ 4.61         \\
		70                   & 741                  & 0.3                     & 0.3                      & 217.0                                      & 4.07 $\pm$ 1.67         & 4.55 $\pm$ 0.77         & 1.52 $\pm$ 0.34         \\
		&                      &                         & 0.5                      & 260.6                                      & 4.56 $\pm$ 0.89         & 12.28 $\pm$ 1.72        & 2.84 $\pm$ 0.46         \\
		&                      &                         & 0.7                      & 248.0                                      & 2.94 $\pm$ 0.37         & 9.72 $\pm$ 2.32         & 1.87 $\pm$ 0.26         \\
		&                      &                         & 1                        & 78.0                                       & 0.07 $\pm$ 0            & 0.05 $\pm$ 0.01         & 0.1 $\pm$ 0.03          \\
		& 1209                 & 0.5                     & 0.3                      & 361.7                                      & 3.27 $\pm$ 0.34         & 0.76 $\pm$ 0.09         & 0.96 $\pm$ 0.1          \\
		&                      &                         & 0.5                      & 460.4                                      & 471.18 $\pm$ 77.27      & 322.99 $\pm$ 112.29     & 324.72 $\pm$ 131.86     \\
		&                      &                         & 0.7                      & 457.7                                      & 308.05 $\pm$ 130.31     & 369.14 $\pm$ 208.88     & 251.21 $\pm$ 96.75      \\
		&                      &                         & 1                        & 382.2                                      & 4.07 $\pm$ 1.08         & 2.93 $\pm$ 1.31         & 20.67 $\pm$ 14.28       \\ \hline
	\end{tabular}
\end{table}

Comparing values of the same column, it can be seen that graphs with a higher density (more edge variables) have a longer solve time. For graphs of a given order and density, we observe the shortest solve times for $m^-/m \in \{0.3, 1\}$ in most cases which are also associated with the two smallest averages of values of $Z^*$. 

\subsection{Convergence of the models with and without the speed-ups}
\begin{figure}[ht]
	\subfloat[The AND model, ER graph]{\includegraphics[width=0.33\textwidth]{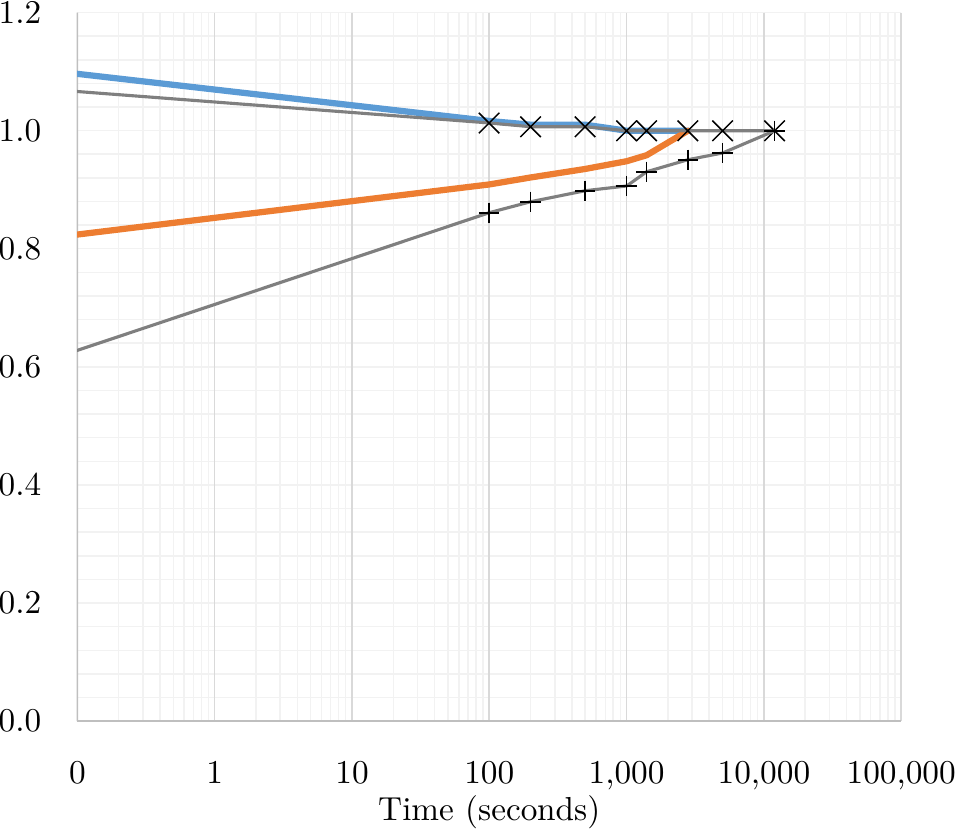}%
		\label{3fig2a}}
	\hfil
	\subfloat[The XOR model, ER graph]{\includegraphics[width=0.33\textwidth]{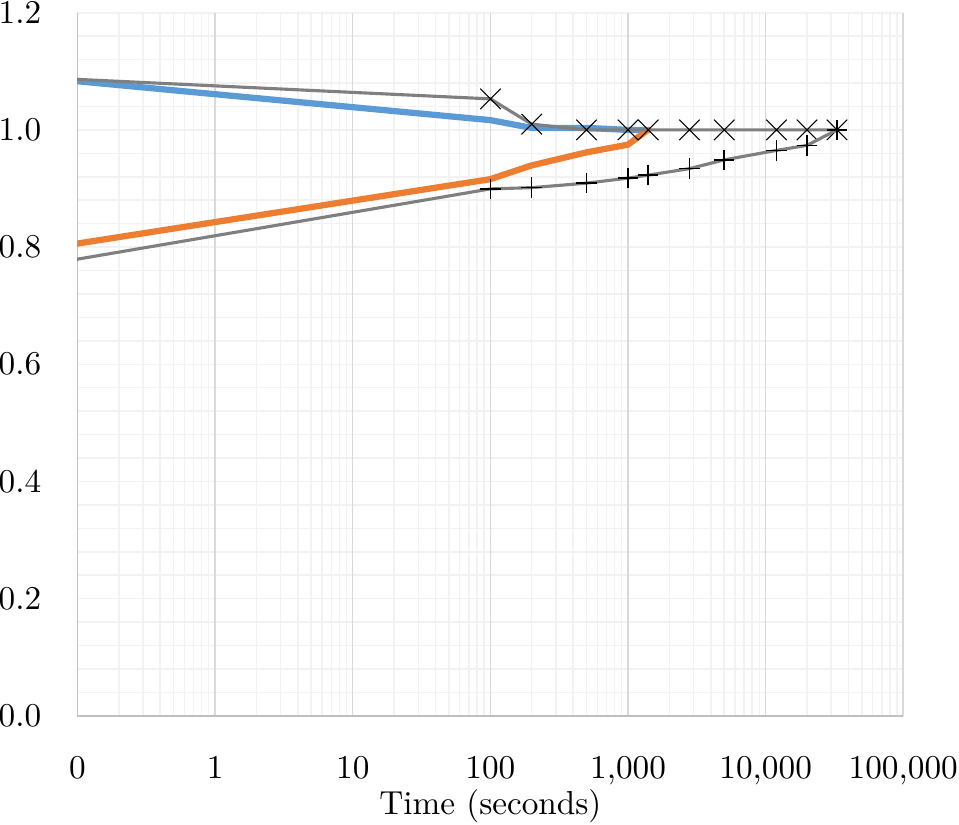}%
		\label{3fig2b}}				 
	\hfil
	\subfloat[The ABS model, ER graph]{\includegraphics[width=0.33\textwidth]{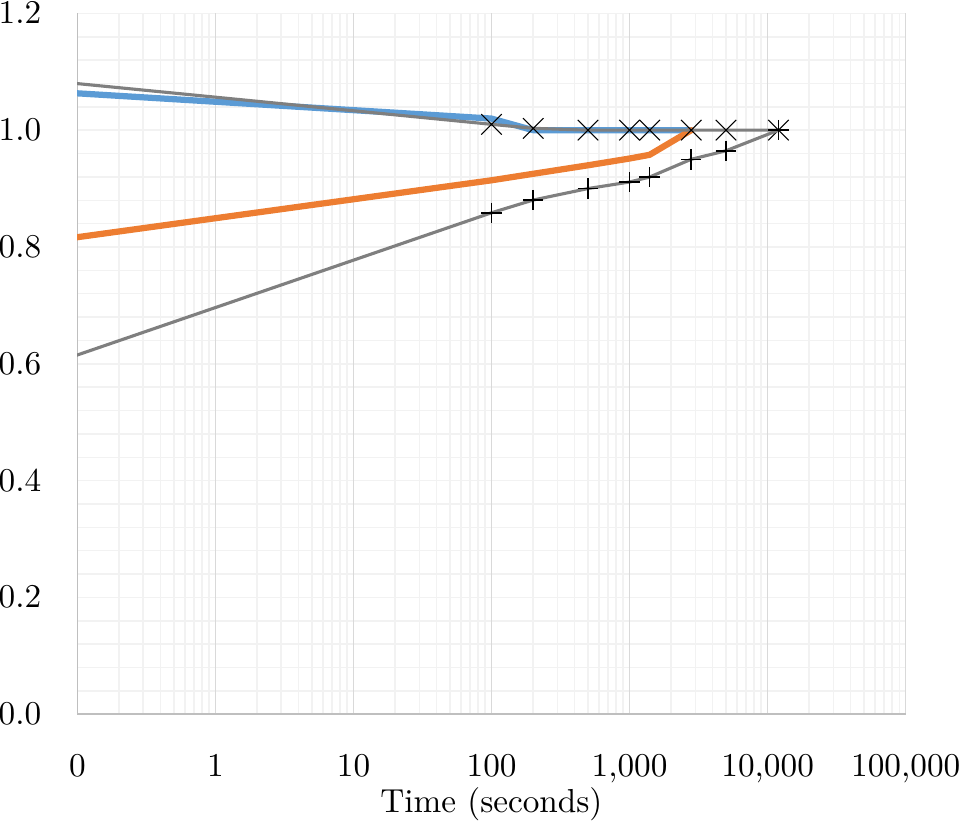}%
		\label{3fig2c}}
	\hfil
	\subfloat[The AND model, BA graph]{\includegraphics[width=0.33\textwidth]{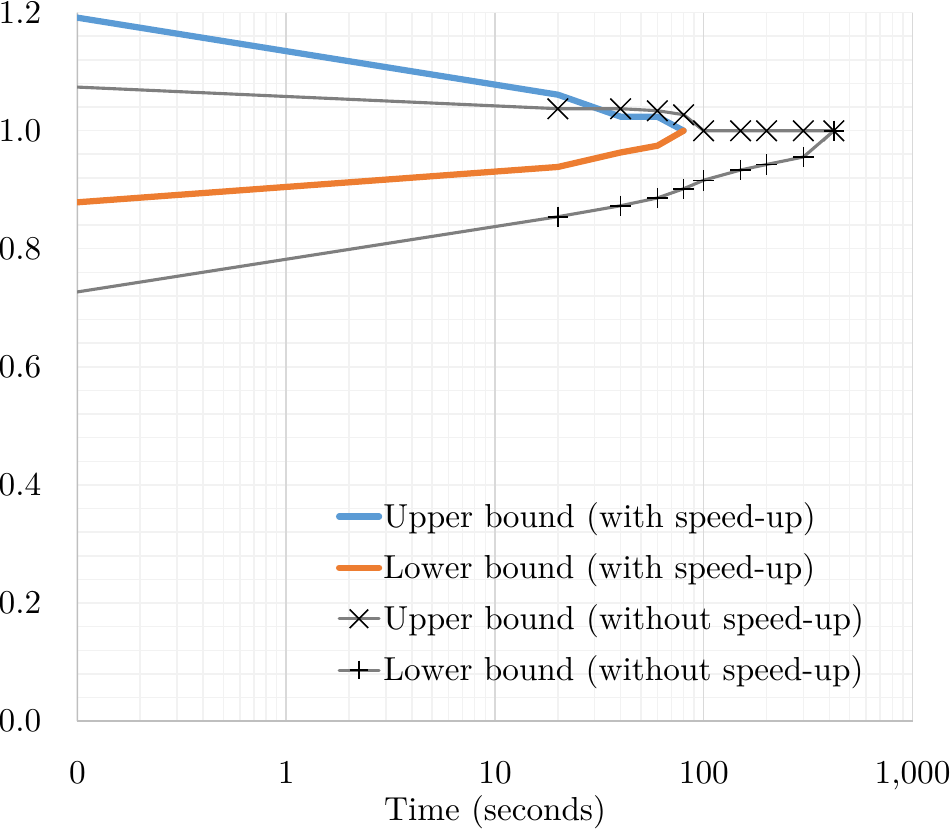}%
		\label{3fig2d}}
	\hfil
	\subfloat[The XOR model, BA graph]{\includegraphics[width=0.33\textwidth]{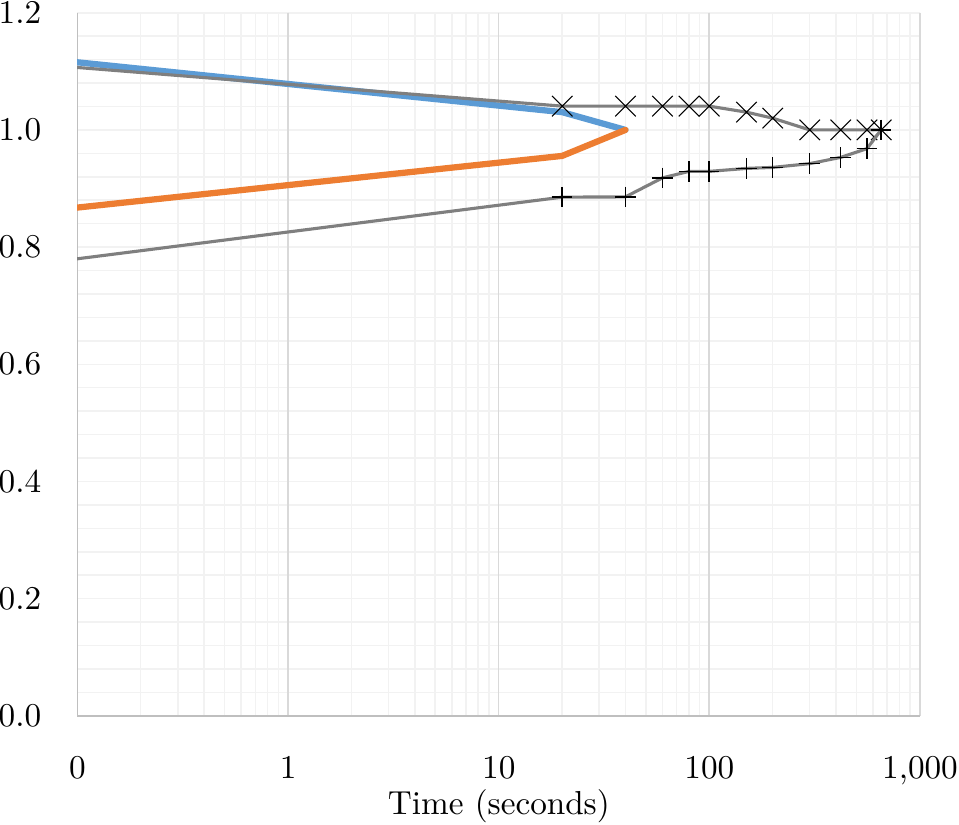}%
		\label{3fig2e}}
	\hfil
	\subfloat[The ABS model, BA graph]{\includegraphics[width=0.33\textwidth]{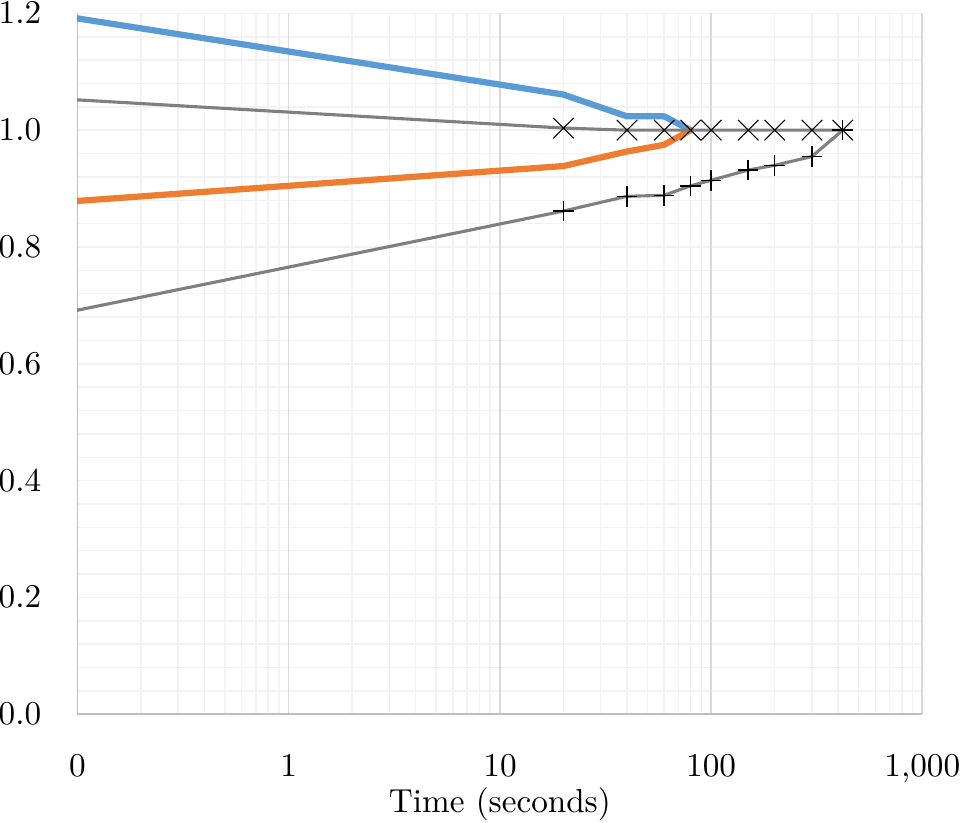}%
		\label{3fig2f}}
	\caption{Normalised upper and lower bounds over time with and without the speed-up techniques for one ER graph and one BA graph with $n=100, m=900, m^-=600$ on a log scale. Vertical axes show normalised upper and lower bounds. (colour version online)}
	\label{3fig2}
\end{figure}

We investigate the algorithm convergence by running the three models with and without the speed-up techniques for one Erd\H{o}s-R\'{e}nyi (ER) random graph and one Barab\'{a}si-Albert (BA) random graph with $n=100, m=900, m^-=600$ and plotting the upper and lower bounds over time. Figure~\ref{3fig2} shows normalised bounds over time on a log scale where the vertical axes represent upper and lower bounds normalised by dividing by the optimal objective function value.

For the randomly generated Erd\H{o}s-R\'{e}nyi graph in Subfigures \ref{3fig2a}, \ref{3fig2b}, and \ref{3fig2c}, the solve times of all three models without the speed-up techniques are over 12000 seconds (and in one case 33000 seconds). These solve times are reduced to less than 2800 seconds (and in one case 1400 seconds) when the speed-up techniques are implemented. 

Subfigures \ref{3fig2d}, \ref{3fig2e}, and \ref{3fig2f} show a considerable solve time improvement for the randomly generated Barab\'{a}si-Albert graph. It takes 420 seconds (80 seconds) for the AND model and the ABS model to find an optimal solution without (with) the speed-up techniques. The XOR model without (with) the speed-up techniques reaches optimality in 655 seconds (40 seconds).
%We provide more extensive comparison of the models equipped with the two speed-up techniques in the next subsection.
\subsection{Largest instances solvable in 10 hours}
Our experiments allow us to discuss the size of the largest graph whose frustration index can be computed in a reasonable time using an extended binary linear model. Two important factors must be taken into consideration in this regard: network properties and processing capacities.
As it is expected from our degree-based prioritised branching in \ref{3ss:branch}, network properties such as degree heterogeneity could have an impact on the solve time. %The results in \ref{3fig2} show Barab\'{a}si-Albert graphs have a shorter solve time compared to Erd\H{o}s-R\'{e}nyi graphs of the same size, order, and proportion of negative edges. 
Moreover, the numerical results in \cite{aref2017computing} suggest that reaching optimality in real signed networks takes a considerably shorter time compared to randomly generated signed networks of comparable size and order, confirming the observations of \cite{dasgupta_algorithmic_2007, huffner_separator-based_2010}. Processing capacities of the computer that runs the optimisation models are also relevant to the size of the largest solvable instance because Gurobi allows using multiple processing cores for exploring the feasible space in parallel \cite{gurobi}. Besides, exploring a large binary tree may require a considerable amount of memory which might be a determining factor in solve time of some instances due to memory limits.

Given a maximum solve time of 10 hours on the current hardware configuration (Intel Core i5 7600 @ 3.50 GHz and 8.00 GB of RAM), random instances with up to 2000 edges were observed to be solvable to global optimality. Regarding real signed graphs which have regularities favouring Gurobi's solver performance, graphs with up to 30000 edges are solvable (to global optimality) within 10 hours. If we use more advanced processing capacities (32 Intel Xeon CPU E5-2698 v3 @ 2.30 GHz processors and 32 GB of RAM), real signed graphs with up to 100000 edges are solvable (to global optimality) within 10 hours \cite{aref2017balance}. 

We have observed in most of our numerical experiments that the branch and bound algorithm finds the globally optimal solution in a fraction of the total solve time, but it takes more time and computations to guarantee the optimality. To give an example, Subfigures \ref{3fig2a}, \ref{3fig2b}, and \ref{3fig2c} show that a considerable proportion of the solve time, ranging in 30\% -- 90\%, is used for guaranteeing optimality after finding the globally optimal solution. One may consider using a non-zero mixed integer programming gap to find solutions within a guaranteed proximity of an optimal solution even if the instance has more than 100000 edges.

\section{Evaluating performance against the literature}\label{3s:evaluate}
In this section, we use both random and real networks to evaluate not only the solve time, but also the output of our models against other methods in the literature.

\subsection{Solve time in random graphs}
In this subsection, we compare the solve time of our algorithm against other algorithms suggested for computing the frustration index. Besides \cite{aref2017computing}, our review of the literature finds only two methods claiming exact computation of the frustration index \cite{brusco_k-balance_2010, huffner_separator-based_2010}. Brusco and Steinley suggested a branch and bound algorithm for minimising the overall frustration (under a different name) for a predefined number of colours \cite{brusco_k-balance_2010}. H\"{u}ffner, Betzler, and Niedermeier have suggested a data-reduction schemes and an iterative compression algorithm for computing the frustration index \cite{huffner_separator-based_2010}.

Brusco and Steinley have reported running times for very small graphs with only up to $n=21$ vertices. While, their exact algorithm fails to solve graphs as large as $n=30$ in a reasonable time \cite{brusco_k-balance_2010}, our binary linear models solve such instances in split seconds.
H\"{u}ffner, Betzler, and Niedermeier have generated random graphs of order $n$ with low densities $(\rho \leq 0.04)$ %by specifying $n$, degree distribution, clustering coefficient, and the percentage of negative edges
to test their algorithm \cite{huffner_separator-based_2010}. The largest of such random graphs solvable by their algorithm in 20 hours has $500$ nodes. They also reported that only 3 out of 5 random graphs with $n \in\{100,200,300,400,500\}$ can be solved by their method in 20 hours. Our three binary linear models solve all such instances in less than 100 seconds.

\subsection{Solve time and algorithm output in real networks} \label{3s:real}
In this section we use signed network data sets from biology and international relations. The frustration index of biological networks has been a subject of interest to measure the distance to \textit{monotonicity} \cite{dasgupta_algorithmic_2007,iacono_determining_2010}. In international relations, the frustration index is used to measure distance to balance for a network of countries \cite{patrick_doreian_structural_2015}. In this section, the frustration index is computed in real biological and international relations networks by solving the three binary linear models coupled with the two speed-up techniques \ref{3ss:branch} -- \ref{3ss:unbalanced}.

We use \textit{effective branching factor} as a performance measure. If the solver explores $b$ branch and bound nodes to find an optimal solution of a model with $v$ variables, the effective branching factor is $\sqrt[v]b$. The most effective branching is obtained when the solver only explores 1 branch and bound node to reach optimality. The effective branching factor for such a case would take value 1 which represents the strength of the mathematical formulation.

% using the Gurobi 7.0.2 Python interface and a desktop computer with an Intel Corei5 4670 @ 3.40 GHz and 8.00 GB of RAM running 64-bit Microsoft Windows 7.
\subsubsection{Biological data sets}
We use the four signed biological networks that were previously analysed by \cite{dasgupta_algorithmic_2007} and \cite{iacono_determining_2010}. The epidermal growth factor receptor (EGFR) pathway \cite{oda2005} is a signed network with 779 edges. The molecular interaction map of a macrophage (macro.) \cite{oda2004molecular} is another well studied signed network containing 1425 edges. We also investigate two gene regulatory networks, related to two organisms: a eukaryote, the \textit{yeast Saccharomyces cerevisiae} (yeast), \cite{Costanzo2001yeast} and a bacterium, \textit{Escherichia coli} (E.coli) \cite{salgado2006ecoli}. The yeast and E.coli networks have 1080 and 3215 edges respectively. The data sets for real networks used in this study are publicly available in a \href{https://figshare.com/articles/Signed_networks_from_sociology_and_political_science_biology_international_relations_finance_and_computational_chemistry/5700832}{Figshare} research data repository \cite{Aref2017data}. For more details on the four biological data sets, one may refer to \cite{iacono_determining_2010}.

We use root node objective, Number of B\&B nodes, effective branching factor, and solve time as performance measures. The performance of three binary linear models can be compared based on these measures in Table~\ref{3tab4} in which values in brackets show the corresponding measure for the case in which speed-up techniques were not used.

\begin{table}[ht]
	%\centering
	\caption{Performance measures for the three binary linear models with (and without) the speed-ups}
	\label{3tab4}
	\begin{tabular}{llllll}
		\hline
		\multicolumn{2}{r}{\begin{tabular}[r]{@{}r@{}}Graph\\ $n, m$\end{tabular}} & \begin{tabular}[c]{@{}l@{}}EGFR\\ 329, 779\end{tabular} & \begin{tabular}[c]{@{}l@{}}Macro.\\ 678, 1425\end{tabular} & \begin{tabular}[c]{@{}l@{}}Yeast\\ 690, 1080\end{tabular} & \begin{tabular}[c]{@{}l@{}}E.coli\\ 1461, 3215\end{tabular} \\ \hline
		{\begin{tabular}[c]{@{}l@{}}Root node objective\end{tabular}}        & {AND} & 28.5            & 67                 & 11.5              & 130.5               \\
		&                      & (13)            & (53)               & (0)               & (4)                 \\
		& {XOR} & 28.5            & 67                 & 11.5              & 130.5               \\
		&                      & (13)            & (53)               & (0)               & (4)                 \\
		& {ABS} & 28.5            & 67                 & 11.5              & 130.5               \\
		&                      & (13)            & (53)               & (0)               & (4)                 \\ \hline
		{\begin{tabular}[c]{@{}l@{}}Number of B\&B nodes\end{tabular}}       & {AND} & 3               & 1                  & 1                 & 31                  \\
		&                      & (91)            & (199)              & (7)               & (279)               \\
		& {XOR} & 1               & 1                  & 1                 & 3                   \\
		&                      & (25)            & (1)                & (1)               & (19)                \\
		& {ABS} & 1               & 1                  & 3                 & 36                  \\
		&                      & (47)            & (456)              & (7)               & (357)               \\ \hline
		{\begin{tabular}[c]{@{}l@{}}Effective branching factor\end{tabular}} & {AND} & 1.0010          & 1                  & 1                 & 1.0007              \\
		&                      & (1.0041)        & (1.0025)           & (1.0011)          & (1.0012)            \\
		& {XOR} & 1               & 1                  & 1                 & 1.0002              \\
		&                      & (1.0029)        & (1)                & (1)               & (1.0006)            \\
		& {ABS} & 1               & 1                  & 1.0004            & 1.0006              \\
		&                      & (1.0027)        & (1.0022)           & (1.0008)          & (1.0010)            \\ \hline
	\end{tabular}
\end{table}

DasGupta et al.\ have suggested approximation algorithms \cite{dasgupta_algorithmic_2007} that are later tested on the four biological networks in \cite{huffner_separator-based_2010}. Their approximation method provides $196 \leq L(G)_{\text{EGFR}} \leq 219$ which our exact model proves to be incorrect. The bounds obtained by implementing their approximation are not incorrect for the other three networks, but they have very large gaps between lower and upper bounds.

H\"{u}ffner, Betzler, and Niedermeier have previously investigated frustration in the four biological networks suggesting a data-reduction schemes and (an attempt at) an exact algorithm \cite{huffner_separator-based_2010}. Their suggested data-reduction schemes can take more than 5 hours for yeast, more than 15 hours for EGFR, and more than 1 day for macrophage if the parameters are not perfectly tuned. Besides the solve time issue, their algorithm provides $L(G)_{\text{EGFR}}=210, L(G)_{\text{macrophage}}=374$, both of which are proven to be incorrect by our results. They report that their algorithm fails to terminate for E.coli \cite{huffner_separator-based_2010}.

Iacono et al.\ have also investigated frustration in the four networks \cite{iacono_determining_2010}. Their heuristic algorithm provides upper and lower bounds for EGFR, macrophage, yeast, and E.coli with 96.37\%, 90.96\%, 100\%, and 98.38\% ratio of lower to upper bound respectively. The comparison of our outputs against those reported in the literature is provided in Table~\ref{3tab5}.

Iacono et al.\ also suggest an upper bound for the frustration index \cite[page 227]{iacono_determining_2010}. However, some values of the frustration index in complete graphs with all negative edges show that their suggested upper bound is incorrect (take a complete graph with 9 nodes and 36 negative edges which has a frustration index of 16 while the bound suggested in \cite{iacono_determining_2010} gives a value of 15). For a more detailed discussion on bounds for the frustration index, one may refer to \cite{aref2017computing,martin2017frustration}.

We also compare our solve times to the best results reported for heuristics and approximation algorithms in the literature. H\"{u}ffner et al.\ have provided solve time results for their suggested algorithm \cite{huffner_separator-based_2010} (if parameters are perfectly tuned for each instance) as well as the algorithm suggested by DasGupta et al.\ \cite{dasgupta_algorithmic_2007}. Iacono et al.\ have only mentioned that their heuristic requires a fairly limited amount of time (a few minutes on an ordinary PC \cite{iacono_determining_2010}) that we conservatively interpret as 60 seconds. 

Table~\ref{3tab5.5} sums up the solve time comparison of our suggested models against the literature in which the values for running our models without the speed-up techniques are provided inside brackets. As the hardware configuration is not reported in \cite{dasgupta_algorithmic_2007, iacono_determining_2010}, we conservatively evaluate the order-of-magnitude improvements in solve time with respect to the differences in computing power in different years.

\begin{table}[ht]
	%\centering
	\caption{Our algorithm output against the best results reported in the literature}
	\label{3tab5}
	\begin{tabular}{p{1.6cm}p{1.7cm}p{1.6cm}p{1.6cm}p{1.3cm}p{1cm}p{1cm}p{1cm}}
		\hline
		
		Author Reference     & DasGupta \quad et~al.\ \cite{dasgupta_algorithmic_2007} & H\"{u}ffner \quad et~al.\ \cite{huffner_separator-based_2010} & Iacono \quad et~al.\ \cite{iacono_determining_2010} & Aref \quad et~al.\ \cite{aref2017computing} &AND \quad \eqref{3eq3.5} & XOR \quad \eqref{3eq4} &ABS \quad \eqref{3eq5}    \\ \hline
		
		EGFR & {[}196, 219{]}$\dagger$                                                           & 210$\dagger$                                                          & {[}186, 193{]}                                                     & 193  & 193 & 193& 193   \\
		Macro. & {[}218,383{]}                                                            & 374$\dagger$                                                          & {[}302, 332{]}                                                     & 332  & 332  & 332  & 332    \\
		Yeast & {[}0, 43{]}                                                              & 41                                                           & 41                                                                & 41   & 41 & 41 & 41    \\
		E.coli & {[}0, 385{]}                                                             & $\ddagger$                                               & {[}365, 371{]}                                                   & 371    & 371  & 371 & 371   \\ \hline
	\end{tabular}
 \\ {$\dagger$ Incorrect results}\\ {$\ddagger$ The algorithm does not converge} 
\end{table}
%\FloatBarrier

\begin{table}[ht]
	%\centering
	\caption{Algorithm solve time in seconds with (and without) the speed-up techniques against the results reported in the literature}
	\label{3tab5.5}
	\begin{tabular}{p{1.5cm}p{1cm}p{1cm}p{1cm}p{1cm}p{1.8cm}p{1.8cm}p{1.8cm}}
		\hline
		Year &2010&2010&2010&2018&2018&2018&2018\\ 
		Reference     & \cite{dasgupta_algorithmic_2007} & \cite{huffner_separator-based_2010} &  \cite{iacono_determining_2010} & \cite{aref2017computing} &AND \eqref{3eq3.5} & XOR \eqref{3eq4} &ABS \eqref{3eq5}    \\ \hline

		EGFR & 420                                                                   & 6480                                                      & \textgreater60                                             & 0.68 & 0.27 (0.82) & 0.21 (0.67)& 0.23 (0.66)\\
		Macro. & 2640                                                                  & 60                                                        & \textgreater60                                                & 1.85 & 0.34 (1.24) & 0.26 (1.37) & 0.49 (1.30) \\
		Yeast & 4620                                                                  & 60                                                        & \textgreater60                                                & 0.33 & 0.18 (0.45) & 0.11 (0.28) & 0.15 (0.39) \\
		E.coli & $\dagger$                                                           & $\ddagger$                                                & \textgreater60                                               &  18.14&  0.99 (1.91) & 1.97 (4.73)  & 0.74 (1.86) \\ \hline   
	\end{tabular}
\\ {$\dagger$ Not reported} \\ {$\ddagger$ The algorithm does not converge}
\end{table}

According to Moore's law \cite{moore1965}, the exponential increase in transistor density on integrated circuits leads to computer power doubling almost every two years. Moore's prediction has been remarkably accurate from 1965 to 2013, while the actual rate of increase in %the maximum number of transistors to fit on an integrated circuit chip
computer power has slowed down since 2013 \cite{mack2015multiple}.

Moore's law ballpark figures allow us to compare computations executed on different hardware in different years. We conservatively estimate a factor of 16 times for the improvements in computer power between 2010 and 2018 to be attributable to hardware improvements.
%Based on Moore's law, we expect 16 times faster computation in 2018 compared to 2010. 
The solve times of the slowest (fastest) model among AND, XOR, and, ABS in Table~\ref{3tab5.5} shows a factor of improvement ranging between $30 - 333$ ($81 - 545$) compared to the fastest solve time in 2010 \cite{dasgupta_algorithmic_2007, huffner_separator-based_2010, iacono_determining_2010}. This shows our solve time improvements are not merely resulted from hardware differences.

While data-reduction schemes \cite{huffner_separator-based_2010} can take up to 1 day for these data sets and heuristic algorithms \cite{iacono_determining_2010} only provide bounds with up to 9\% gap from optimality, our three binary linear models equipped with the speed-up techniques (\ref{3ss:branch} -- \ref{3ss:unbalanced}) solve the four instances to optimality in a few seconds. 

\subsubsection{International relations data sets}

We also compute the frustration index for two data sets of international relations networks. In international relation networks, countries and their relations are represented by nodes and edges of signed graphs. We use the Correlates of War (CoW) \cite{correlatesofwar2004} data set which has 51 instances of networks with up to 1247 edges \cite{patrick_doreian_structural_2015} and the United Nations General Assembly (UNGA) \cite{macon2012community} data set which has 62 instances with up to 15531 edges when converted into signed networks by \cite{figueiredo2014maximum}. Figueiredo and Frota provide detailed explanation on the process of creating signed networks from the UNGA data \cite{figueiredo2014maximum}. %The Correlates of War data set contains 51 time windows of a temporal network representing signed international relations among countries starting with 1946-1949 time window and ending with 1996-1999 time window \cite{correlatesofwar2004}. Both size and order change in each time window.
%The first time window of the temporal network has 362 edges while the last time window contains 1247 edges.

The CoW signed network data set is constructed by Doreian and Mrvar \cite{patrick_doreian_structural_2015} based on signed international relations between countries in 1946-1999. In their analysis, some numerical results provided on the CoW data set are referred to as line index \cite{patrick_doreian_structural_2015}. However, the values of $L(G)$ we have obtained using our optimisation models prove that values reported in \cite{patrick_doreian_structural_2015} for the 51 time frames of the network are never the smallest number of edges whose removal results in balance. Doreian and Mrvar have not reported any solve time, but have suggested that determining their line index is in general a polynomial-time hard problem \cite{patrick_doreian_structural_2015}. The solve times of our models for each instance of the CoW data set is $\leq 0.1$ seconds.

\begin{figure}[ht]
	\includegraphics[width=1\textwidth]{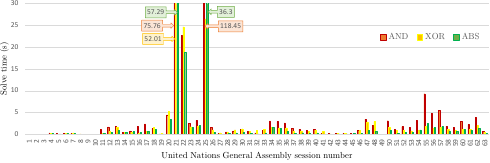}%
	\caption{Solve times of AND, XOR, and ABS models tested on the UNGA instances (colour version online)}
	\label{3fig2.5}
\end{figure}

We also tested our three models on the UNGA instances. The UNGA data set is based on voting on the UN resolutions. In this data set, instances refer to annual UNGA sessions between 1946 and 2008. Figure~\ref{3fig2.5} shows the solve times of our models for instances of this data set. 

As it can be seen in Figure~\ref{3fig2.5}, most UNGA instances can be solved in less that 5 seconds using any of the three models. The XOR and the ABS models solve all UNGA instances in less than a minute, while solving the AND model for instance 21 and instance 25 takes about 75 and 118 seconds respectively. These two harder instances have the highest values of the frustration index ($L(G)=616$ and $L(G)=611$ respectively) in the UNGA data set. 

\section{Other formulations and extensions} \label{3s:future}
In this section we provide an alternative formulation and two extensions to the 2-colour minimum frustration count optimisation problem.

\subsection{Max (2,2)-CSP formulation and theoretical results} \label{ss:maxCSP}
In this subsection, we formulate the problem of computing the frustration index as a constraint satisfaction problem in \eqref{3eq12} and provide theoretical results on the fastest known algorithms. Computation of the frustration index can be formulated as a Maximum 2-Constraint Satisfaction Problem with 2 states per variable (Max (2,2)-CSP) with $n$ variables and $m$ constraints.

The signed graph, $G(V,E,\sigma)$, is the input constraint graph. We consider a score for each edge $(i,j)$ depending on its sign $\sigma_{ij}$ and the assignment of binary values to its endpoints. In the formulation provided in \eqref{3eq12}, the dyadic score function $S_{(i,j)}:\{0,1\}^2 \rightarrow \{0,1\}$ determines the satisfaction of edge $(i,j)$ accordingly (score $1$ for satisfied and score $0$ for frustrated). The output of solving this problem is the colouring function $\phi: V \rightarrow \{0,1\}$ which maximises the total number of satisfied edges as score function $S(\phi)$. 

Denoting the maximum score function value by $S^*(\phi)$, the frustration index can be calculated as the number of edges that are not satisfied $m-S^*(\phi)$.

\begin{equation} \label{3eq12}
\begin{split}
\max_\phi S(\phi)=\sum_{(i,j) \in E} S_{(\phi(i), \phi(j))} \\
%\max_\phi S(\phi)=\sum_{(u,v) \in E^-} S^-_{(\phi(u),\phi(v))} + \sum_{(u,v) \in E^+} S^+_{(\phi(u),\phi(v))} \\ 
S_{(i,j)}=\{ ((0,0),(1+\sigma_{ij})/2),\\
((0,1),(1-\sigma_{ij})/2),\\
((1,0),(1-\sigma_{ij})/2),\\
((1,1),(1+\sigma_{ij})/2)\} 
%S^-_{(u,v)}=\{ ((0,0),0), ((0,1),1),((1,0),1),((1,1),0)\} \\
%S^+_{(u,v)}=\{ ((0,0),1), ((0,1),0),((1,0),0),((1,1),1)\}
\end{split}
\end{equation}

According to worst-case analyses, the fastest known algorithm \cite{koivisto2006optimal} with respect to $n$ solves Max (2,2)-CSP in $\mathcal{O}(nm2^{n\omega/3})$, where $\omega$ is the matrix multiplication exponent. Since $\omega < 2.373$ \cite{LeGall2014}, the running time of the algorithm from \cite{koivisto2006optimal} is $\mathcal{O}(1.7303^n)$. It improves on the previous fastest algorithm \cite{WILLIAMS2005} only in the polynomial factor of the running time. With respect to $n$, the algorithm in \cite{koivisto2006optimal} is the fastest known algorithm for MAXCUT, and therefore for computing the frustration index as well. Both algorithms \cite{koivisto2006optimal,WILLIAMS2005} use exponential space and it is open whether MAXCUT can be solved in $\mathcal{O}(c^n)$ for some $c<2$ when only polynomial space is allowed.

With respect to the number $m$ of edges, the Max (2,2)-CSP formulation in \eqref{3eq12} enables the use of algorithms from \cite{Gaspers2017} and \cite{SCOTT2007}. The first algorithm uses $2^{(9m/50 + \mathcal{O}(m))}$ time and polynomial space \cite{Gaspers2017}, while the second algorithm uses $2^{(13m/75 + \mathcal{O}(m))}$ time and exponential space. With respect to the number of edges, these two algorithms \cite{Gaspers2017,SCOTT2007} are also the fastest algorithms known for MAXCUT, and therefore for computing the frustration index.

\subsection{Weighted minimum frustration count optimisation problem}
We extend the 2-colour minimum frustration count optimisation problem for a graph with weights $w_{ij} \in [-1,1]$ instead of the signs $a_{ij} \in \{-1,1\}$ on the edges. We call such a graph a \textit{weighted signed graph}.

Taking insights from \eqref{3eq6}, the frustration of edge $(i,j) \in E$ with weight $w_{ij}$ can be represented by $f_{ij}={(1-w_{ij})}/{2} + w_{ij}(x_i +x_j - 2x_{ij}) $ using the binary variables $x_i , x_j, x_{ij}$ of the AND model \eqref{3eq3.5}. Note that, the frustration of an edge in a weighted signed graph is a continuous variable in the unit interval $f_{ij} \in [0,1]$.

Note that, $a_{ij}x_{ij} \leq (3a_{ij}-1)(x_i + x_j)/4 + {(1-a_{ij})}/{2}$ embodies all constraints for edge $(i,j)$ in the AND model regardless of the edge sign. Accordingly, the constraints of the AND model can be modified to incorporate weights $w_{ij}$. The weighted minimum frustration count optimisation problem can be formulated as a binary linear programming model in \eqref{3eq9}. 
\begin{equation}\label{3eq9}
\begin{split}
\min_{x_i: i \in V, x_{ij}: (i,j) \in E} Z &= \sum\limits_{(i,j) \in E} {(1-w_{ij})}/{2} + w_{ij}(x_i +x_j - 2x_{ij})\\
\text{s.t.} \quad
w_{ij}x_{ij} &\leq (3w_{ij}-1)(x_i + x_j)/4 + {(1-w_{ij})}/{2} \quad \forall (i,j) \in E\\
x_{i} &\in \{0,1\} \quad  \forall i \in V \\
x_{ij} &\in \{0,1\} \quad \forall (i,j) \in E 
\end{split}
\end{equation}

We have generated random weighted signed graphs to test the model in \eqref{3eq9}. Our preliminary results show that the weighted version of the problem \eqref{3eq9} is solved faster than the original models for signed graphs.

\subsection{Multi-colour minimum frustration count optimisation problem}

\begin{figure}[ht]
	\subfloat[An example graph with $n=4,$ $m^-=4,$ $ m^+=1$]{\includegraphics[height=1in]{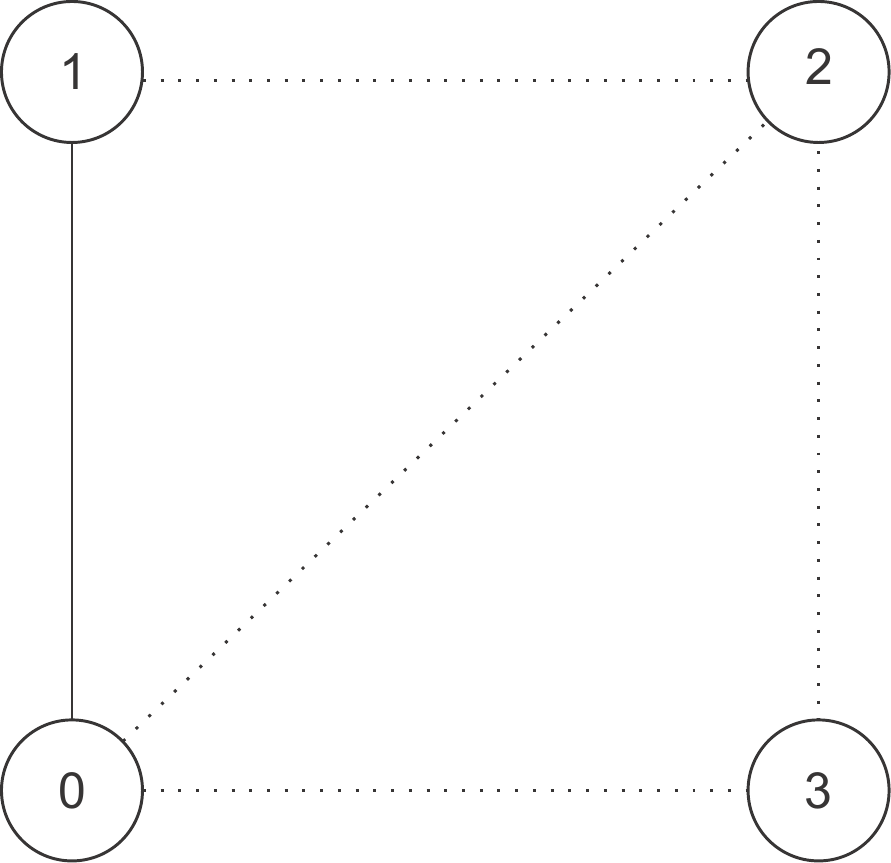}%
		\label{3fig3a}} 
	\hfil
	\subfloat[One colour resulting in four frustrated edges]{\includegraphics[height=1in]{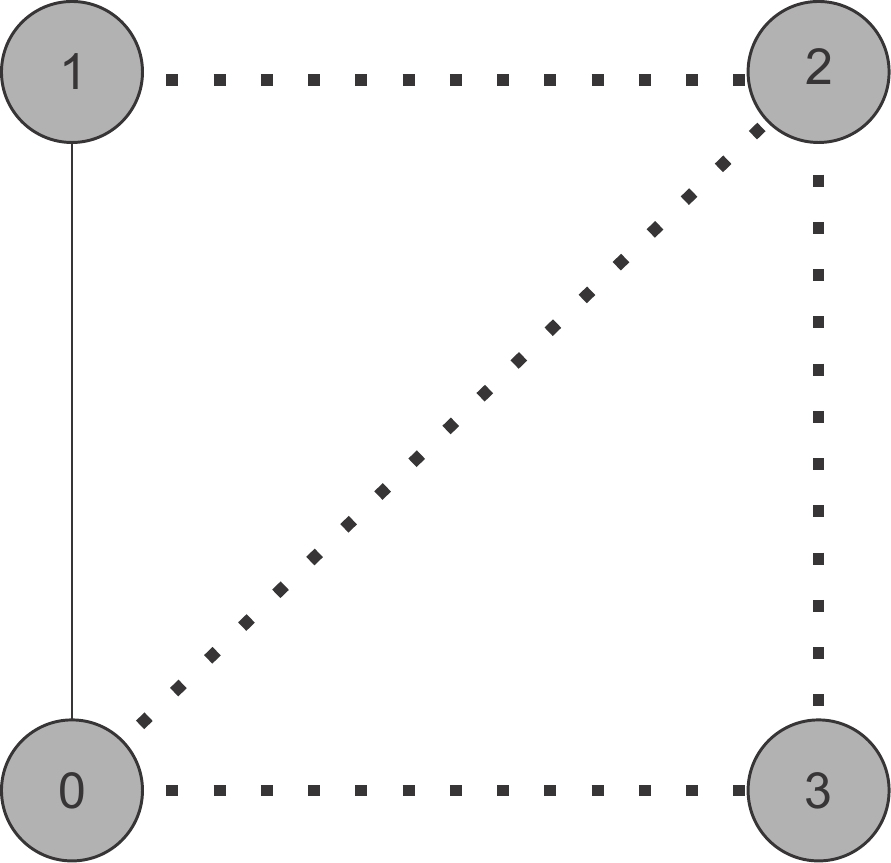}%
		\label{3fig3b}}
	\hfil
	\subfloat[Two colours resulting in one frustrated edge]{\includegraphics[height=1in]{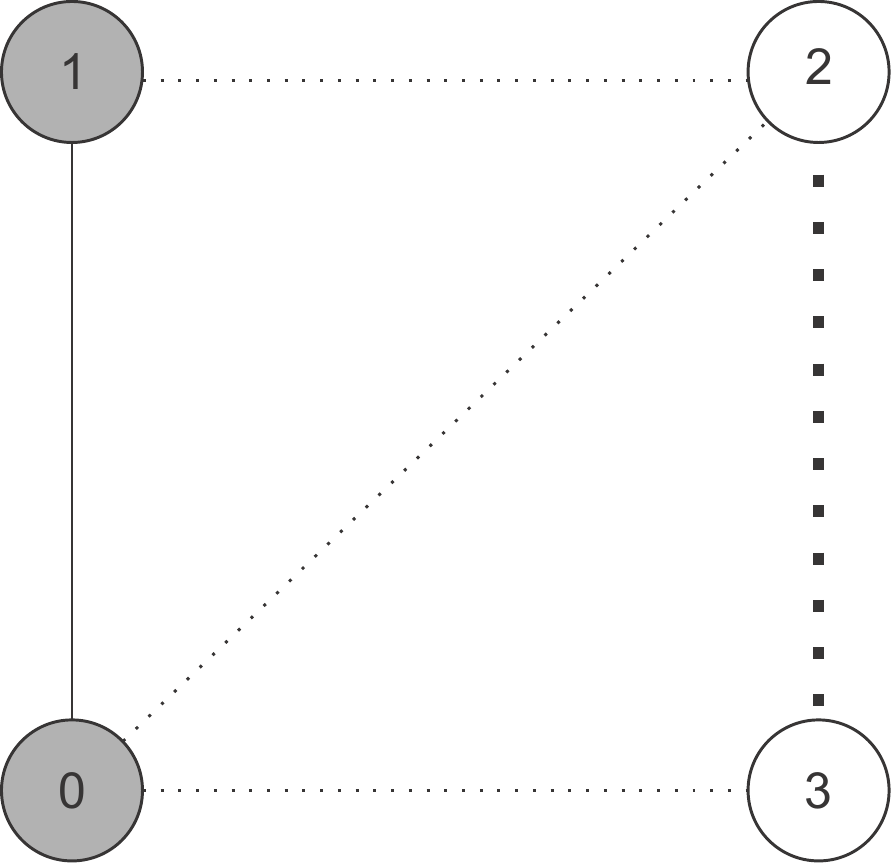}%
		\label{3fig3c}}
	\hfil
	\subfloat[Three colours resulting in no frustrated edge]{\includegraphics[height=1in]{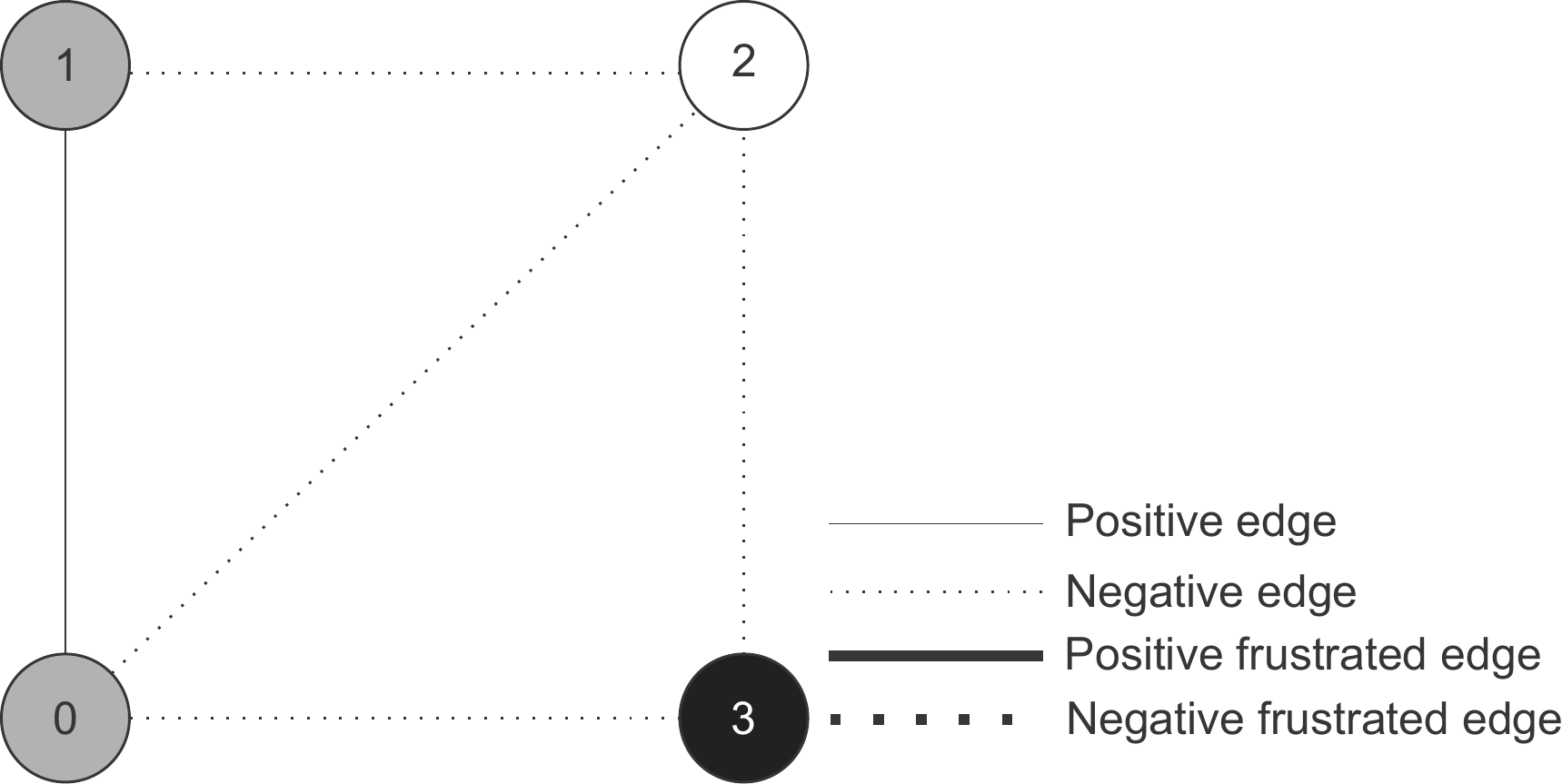}%
		\label{3fig3d}}
	\caption{The frustrated edges represented by dashed lines for the multi-colour minimum frustration count optimisation problem.}% The figure is produced using Adobe Illustrator.}
	\label{3fig3}
\end{figure}

We formulate another extension to the 2-colour minimum frustration count optimisation problem by allowing more than 2 colours to be used. As previously mentioned in Subsection~\ref{3s:related}, a signed network is $k$-balanced if and only if its vertex set can be partitioned into $k$ subsets (for some fixed $k\geq 1$) such that each negative edge joins vertices belonging to different subsets \cite{Davis}. Figure~\ref{3fig3} demonstrates an example graph and the frustrated edges for various numbers of colours. Subfigure~\ref{3fig3d} shows that the graph is weakly balanced. 

The harder problem of finding the minimum number of frustrated edges where $k$ is not specified in advance (an arbitrary number of node colours) is referred to as the Correlation Clustering problem. As mentioned in Subsection~\ref{3s:related}, another integer linear programming formulation for the correlation clustering problem is suggested by \cite{demaine2006correlation} which is widely used in the literature \cite{figueiredo2013mixed, drummond2013efficient, levorato2015ils}.

In the multi-colour minimum frustration count optimisation problem, each node may be given one of a set of colours $C=\{1, 2, 3,..., k:=|C|\}$. Assume $c_i \in C$ is the colour of node $i$. We consider that a positive edge $(i,j) \in E^+$ is frustrated (indicated by $f_{ij}=1$) if its endpoints $i$ and $j$ are coloured differently, i.e.,\ $c_i \ne c_j$; otherwise it is not frustrated (indicated by $f_{ij}=0$). A negative edge $(i,j) \in E^-$ is frustrated (indicated by $f_{ij}=1$) if $c_i = c_j$; otherwise it is not frustrated (indicated by $f_{ij}=0$). %This gives rise to the optimisation problem in Eq.\ \eqref{3eq10}.
%\begin{equation}\label{3eq10}
%\min_{(c_1, c_2, ..., c_{|V|})  \in C^{|V|}} \sum_{(i,j) \in E}  f_{ij}
%\end{equation}

Using binary variables $x_{ic}=1$ if node $i \in V$ has colour $c\in C$ (and $x_{ic}=0$ otherwise), we formulate this as the following binary linear model in Eq.\ \eqref{3eq11}.
\begin{equation} \label{3eq11}
\begin{split}
\min \sum_{(i,j) \in E}   f_{ij}  \\
\text{s.t.} \quad \sum_{c \in C} x_{ic} &= 1 \quad \forall i \in V \\
f_{ij}  &\ge  x_{ic} - x_{jc} \quad \forall (i,j) \in E^+,  ~\forall c \in C \\
f_{ij}  &\ge  x_{ic} + x_{jc} -1 \quad \forall (i,j) \in E^-,  ~\forall c \in C \\
x_{ic} &\in \{0,1\} \quad  \forall i \in V,  ~\forall c \in C \\
f_{ij} &\in \{0,1\} \quad \forall (i,j) \in E 
\end{split}
\end{equation}

If we have just two colours, then we use $x_{i} \in \{0,1\}$ to denote the colour of node $i$. This gives the XOR model expressed in Eq.\ \eqref{3eq4}.

Solving the problem in \eqref{3eq11} provides us with the minimum number of frustrated edges in the $k$-colour setting. This number determines how many edges should be removed to make the network $k$-balanced. For a more general formulation of partitioning graph vertices into $k$ sets, one may refer to \cite{ales2016polyhedral} where numerical results for graphs with up to $20$ nodes are provided. 

\section{Conclusion} \label{3s:conclu}

In this study, we provided an efficient method for computing a standard measure in signed graphs which has many applications in different disciplines. The present study suggested efficient mathematical programming models and speed-up techniques for computing the frustration index in graphs with up to 15000 edges on inexpensive hardware.

We developed three new binary optimisation models which outperform previous methods by large factors. We also suggested prioritised branching and valid inequalities which make the binary linear optimisation models several times (see Table \ref{3tab5.5}) faster than recently developed models \cite{aref2017computing} and capable of processing relatively large instances.

Extensive numerical results on random and real networks were provided to evaluate computational performance and underline the superiority of our models to other methods in the literature in both solve time and algorithm output. We also formulated the problem as a constraint satisfaction model and provided theoretical results on the fastest known algorithms for computing the frustration index with respect to the number of nodes and the number of edges. We also provided two extensions to the model for future investigation. 

%Due to the connection between other signed graph problems like signed graph embedding and balance \cite{Pardo2015}, one may investigate the role of the frustration index in such areas.
%Considering the recent developments in the area of quantum computing and the ever-increasing size of practical instances of the problem, another possibility for future investigation is applying more advanced computing technology for solving the UBQP formulation of the problem or analysing the boolean quadric polytope using a polyhedral approach.
%the connection between balance and linear programming models with unimodular constraint matrix \cite{zaslavsky1982signed}
%, to topics of interest in signed networks including clustering \cite{Gallier16} and opinion dynamics \cite{li_voter_2015}.

%From a practical viewpoint, international relationships is a crucial area to implement signed network structural analysis. Having an efficient method of computing frustration index in hand, signed international relations data set can be investigated in terms of minimal deletion sets of relations that need to be protected to avoid cold war bi-polar structures.
%The literature in international relations offers data on formal alliances between some countries as well as hostile relations between some others. Middle East is an interesting region due to the number of conflicts among states of a region. A signed graph can be readily designed where balance theory and frustration index predict which relations need to be protected to avoid a cold war bi-polarity situation in the region.

\section*{Acknowledgements}
The authors thank the anonymous referees, Serge Gaspers, Gregory Gutin, and Jeffrey Linderoth for valuable comments, Yuri Frota for providing data on United Nations General Assembly instances, and Serge Gaspers for his contribution to Subsection \ref{ss:maxCSP}.

\section*{ORCID}
\noindent
\textit{Samin Aref} \quad \href{http://orcid.org/0000-0002-5870-9253}{http://orcid.org/0000-0002-5870-9253} \\
\noindent
\textit{Andrew J. Mason} \quad \href{http://orcid.org/0000-0001-9848-6595}{http://orcid.org/0000-0001-9848-6595} \\
\noindent
\textit{Mark C. Wilson} \quad \href{http://orcid.org/0000-0002-3343-7458}{http://orcid.org/0000-0002-3343-7458}

\bibliography{refs}
\bibliographystyle{acm}

\end{document}